\documentclass[12pt]{article} 
\usepackage[sectionbib]{natbib}
\usepackage{array,epsfig,fancyheadings,rotating,wrapfig,float}
\usepackage[]{hyperref}  
\usepackage{xcolor}
\usepackage{amsfonts}
\usepackage{multirow}
\usepackage{soul}

\usepackage{sectsty, secdot}
\sectionfont{\fontsize{12}{14pt plus.8pt minus .6pt}\selectfont}
\renewcommand{\theequation}{\thesection\arabic{equation}}
\subsectionfont{\fontsize{12}{14pt plus.8pt minus .6pt}\selectfont}

\usepackage{amsmath}
\usepackage{amssymb}
\usepackage{amsfonts}
\usepackage{multirow}
\usepackage{amsthm}

\newcommand{\pcite}[1]{\citeauthor{#1}'s \citeyearpar{#1}}

\graphicspath{{./plots/}}

\newtheorem{theorem}{Theorem}

\theoremstyle{definition}

\newcommand{\ind}{\buildrel \text{ind} \over \sim}
\newcommand{\Real}{\mathbb{R}}
\newcommand{\cas}{\buildrel \text{a.s.} \over \longrightarrow}

\newcommand{\cd}{\buildrel d \over \rightarrow}
\newcommand{\sX}{\textsf{X}}

\newcommand{\ba}{\boldsymbol{a}}

\newcommand{\bd}{\boldsymbol{d}}

\newcommand{\bmu}{\boldsymbol{\mu}}
\newcommand{\hatbd}{\boldsymbol{\hat{d}}}

\newcommand{\bzeta}{\boldsymbol{\zeta}}
\newcommand{\hatbzeta}{\boldsymbol{\hat{\zeta}}}

\newcommand{\bz}{\boldsymbol{z}}
\newcommand{\by}{\boldsymbol{y}}
\newcommand{\bx}{\boldsymbol{x}}
\newcommand{\bq}{\boldsymbol{q}}
\newcommand{\bpi}{\boldsymbol{\pi}}

\newcommand{\bxi}{{\boldsymbol{\xi}}}
\newcommand{\bbeta}{{\boldsymbol{\eta}}}

\newcount\dotcnt\newdimen\deltay
\def\Ddot#1#2(#3,#4,#5,#6){\deltay=#6\setbox1=\hbox to0pt{\smash{\dotcnt=1
\kern#3\loop\raise\dotcnt\deltay\hbox to0pt{\hss#2}\kern#5\ifnum\dotcnt<#1
\advance\dotcnt 1\repeat}\hss}\setbox2=\vtop{\box1}\ht2=#4\box2}

\DeclareMathOperator{\Cov}{Cov}
\DeclareMathOperator*{\argmax}{argmax}

\begin{document}


\renewcommand{\baselinestretch}{2}
\renewcommand{\floatpagefraction}{.9}

\markright{ \hbox{\footnotesize\rm Statistica Sinica
}\hfill\\[-13pt]
\hbox{\footnotesize\rm
}\hfill }

\markboth{\hfill{\footnotesize\rm VIVEKANANDA ROY AND EVANGELOS EVANGELOU} \hfill}
{\hfill {\footnotesize\rm SELECTING PROPOSALS FOR IMPORTANCE SAMPLING} \hfill}

\renewcommand{\thefootnote}{}
$\ $\par


\fontsize{12}{14pt plus.8pt minus .6pt}\selectfont \vspace{0.8pc}
\centerline{\large\bf SELECTION OF PROPOSAL DISTRIBUTIONS FOR}
\vspace{2pt} 
\centerline{\large\bf MULTIPLE IMPORTANCE SAMPLING}
\vspace{.4cm} 
\centerline{Vivekananda Roy and Evangelos Evangelou} 
\vspace{.4cm} 
\centerline{\it Iowa State
  University, USA and University of
    Bath, UK}
 \vspace{.55cm} \fontsize{9}{11.5pt plus.8pt minus.6pt}\selectfont


\begin{quotation}
\noindent {\it Abstract:}
The naive importance sampling (IS) estimator generally does not
work well in examples involving simultaneous inference on several
targets, as the importance weights can take arbitrarily large values,
making the estimator highly unstable. In such situations, alternative
multiple IS estimators involving samples from multiple proposal
distributions are preferred. Just like the naive IS, the success of
these multiple IS estimators crucially depends on the choice of the
proposal distributions. The selection of these proposal distributions
is the focus of this article. We propose three methods: (i)~a
geometric space filling approach, (ii)~a minimax variance
approach, and (iii)~a maximum entropy approach.  The
first two methods are applicable to any IS estimator, whereas the
third approach is described in the context of \pcite{doss:2010}
two-stage IS estimator. For the first method, we propose a suitable
measure of `closeness' based on the symmetric Kullback-Leibler
divergence, while the second and third approaches use estimates of
asymptotic variances of \pcite{doss:2010} IS estimator and
\pcite{geye:1994} reverse logistic regression estimator, respectively. Thus, when
samples from the proposal distributions are obtained by running Markov
chains, we provide consistent spectral variance estimators for these
asymptotic variances. The proposed methods for selecting proposal
densities are illustrated using various detailed examples.
\vspace{9pt}

\noindent {\it Key words and phrases:}
Bayes factor, central limit theorem, Markov chain, marginal likelihood,
polynomial ergodicity, reverse logistic regression.
\par
\end{quotation}\par

\def\thefigure{\arabic{figure}}
\def\thetable{\arabic{table}}

\renewcommand{\theequation}{\thesection.\arabic{equation}}

\fontsize{12}{14pt plus.8pt minus .6pt}\selectfont

\section{Introduction}
\label{sec:int}
Importance sampling (IS) is a popular Monte Carlo procedure where
samples from one distribution are weighted to estimate features of
other distributions. Here, we consider IS in the context of the
following problem. Let $\Pi$ be the family of target densities on the
space $\sX$ with respect to a measure $\mu$ where
$\pi(x) = \nu(x)/\theta \in \Pi$. Here, $\nu(x)$ is known, but the
normalizing constant $\theta=\int_{\sX} \nu(x) \mu(dx)$ is
unknown. Let $f$ be a $\pi$-integrable, real-valued function defined
on $\sX$ for all $\pi \in \Pi$. There are two goals. The first goal is
to estimate the normalizing constants $\theta$ up to a constant
of proportionality for all $\pi \in \Pi$. The second goal is to
estimate the integrals $E_\pi f := \int_{\sX} f(x) \pi(x) \mu(dx)$ for
all $\pi \in \Pi$. Estimation of normalizing constants plays an
important role in both frequentist and Bayesian inference, as well as
in other areas, like statistical physics. In Bayesian statistics, the
ratio of normalizing constants for two different posteriors is the
Bayes factor, which is at the core of Bayesian hypothesis testing and
model selection \citep{doss:2010}. The empirical Bayes estimate
corresponds to the value of a hyper parameter where the normalizing
constant (marginal likelihood) attains its maximum \citep{doss:2010,
  roy:evan:zhu:2016}. In latent variable models e.g. generalized
linear mixed models, the ratio of the normalizing constants is the
likelihood ratio for hypothesis testing \citep{chri:2004}. The
normalizing constants also need to be estimated in the problems
involving intractable likelihoods, e.g., exponential random graph
models and autologistic models \citep{geye:thomp:1992}. Similarly, in
statistical physics, an important problem is the estimation of some
normalizing constants known as the partition function. On the other
hand, estimation of (posterior) means of certain functions $f$ as the
posterior density varies is the key issue of Bayesian sensitivity
analysis \citep{buta:doss:2011}. In Bayesian penalized regression
methods, plotting regularization paths boils down to estimating means
of regression coefficients as the penalty parameters vary
\citep{roy:chak:2017}.

The two objectives mentioned above can be accomplished using naive importance
sampling.  Let $q_1(x) = \varphi_1(x)/c_1$ be another density
on $\sX$ with respect to $\mu$ such that we are able to generate samples from $q_1$, and $\nu(x) =0$ whenever
$\varphi_1(x) =0$. Indeed, if $\{X_i\}_{i=1}^{n}$ is either independent and identically distributed (iid) samples from
$q_1$ or a positive Harris recurrent Markov chain with invariant density $q_1$,
then the naive IS estimator is consistent, that is,
\begin{equation}
  \label{eq:stis}
  \frac{1}{n} \sum_{i=1}^{n} \frac{\nu(X_i)}{\varphi_1(X_i)} \cas \int_{\sX}  \frac{\nu(x)}{\varphi_1(x)} q_1(x) \,\mu(dx)= \frac{\theta}{c_1}
  \int_{\sX}  \frac{\nu(x)/\theta}{\varphi_1(x)/c_1} q_1(x) \,\mu(dx) = \frac{\theta}{c_1}.
\end{equation}
Similarly, $E_\pi f$ can be estimated by the ratio of
$(1/n) \sum_{i=1}^{n} [f(X_i) \nu(X_i)/\varphi_1(X_i)]$ and the
estimator in \eqref{eq:stis}.  These naive IS estimators suffer from
high variance when the target probability density function (pdf) $\pi$
is not `close' to the proposal pdf $q_1$ \citep{geye:2011b} because,
in that case, the ratio $\nu(X_i)/ \varphi_1(X_i)$ takes arbitrarily large values
for some samples $X_i$'s.

To alleviate this issue, samples from multiple proposals, properly
weighted, can be used, as done in the variants of multiple importance
sampling \citep{veac:guib:1995, owen:zhou:2000,
  elvi:mart:luen:buga:2019}, umbrella sampling \citep{geye:2011b,
  doss:2010}, parallel, serial or simulated tempering
\citep{geor:doss:2017, geye:thomp:1995, mari:pari:1992}. In IS
estimation based on multiple proposal densities, the single density
$q_1$ is generally replaced with a linear combination of $k$ densities
\citep{geye:2011b}. In particular, let $q_i(x) = \varphi_i(x)/c_i$,
for $i=1, \dots, k$, be $k$ densities from the set of potential
proposal densities $Q \equiv \{q(x) = \varphi(x)/c\}$, where the
$\varphi_i$'s are known but the $c_i$'s may be unknown. Let
$\ba= (a_1, \dots, a_k)$ be a vector of $k$ positive constants such
that $\sum_{i=1}^k a_i=1$, $\overline{q} \equiv \sum_{i=1}^k a_i q_i$,
$d_i = c_i/c_1$ for $i=1,2, \dots, k$ with $d_1 = 1$, and
$\bd \equiv (c_2 / c_1, \ldots, c_k / c_1)$. For $l=1, \dots, k$, let
$\{X_i^{(l)}\}_{i=1}^{n_l}$ be either iid samples from $q_l$ or a
positive Harris recurrent Markov chain with invariant density $q_l$.
Then as $n_l \rightarrow \infty, \forall \;
l$, 
\begin{align}
  \label{eq:mainest}
  \hat{u} \equiv  \sum_{l=1}^k \frac{a_l} {n_l} \sum_{i=1}^{n_l}
            \frac{\nu(X_i^{(l)})} {\sum_{j=1}^k a_j \varphi_j(X_i^{(l)}) / d_j}
   & \cas  \sum_{l=1}^k a_l \int_{\sX}
       \frac{\nu(x)} {\sum_{j=1}^k a_j \varphi_j(x) / d_j} q_l(x)
       \, \mu(dx) \\ &= \frac{1}{c_1} \int_{\sX} \frac{\nu(x)} {\overline{q}(x)} \overline{q}(x) \, \mu(dx) = \frac{\theta}{c_1} \nonumber.
\end{align}
Similarly, $E_\pi f$ is estimated by
$\hat{\eta}^{[f]} \equiv \hat{v}^{[f]}/\hat{u}$ where
\[
\hat{v}^{[f]} := \sum_{l=1}^k \frac{a_l} {n_l} \sum_{i=1}^{n_l}  \frac{f(X_i^{(l)}) \nu(X_i^{(l)})} {\sum_{j=1}^k a_j \varphi_j(X_i^{(l)}) / d_j}.
\]
Estimation using \eqref{eq:mainest} has been considered in several
articles \citep[see, e.g.][]{ gill:vard:well:1988,
  kong:mccu:meng:nicol:tan:2003, meng:wong:1996, tan:2004,
  vard:1985,buta:doss:2011, geye:1994, tan:doss:hobe:2015}. There are
alternative weighting schemes proposed in the literature, e.g., the
population Monte Carlo of \citet{capp:guil:mari:robe:2004}, although
none is as widely applicable as~\eqref{eq:mainest}. If the normalizing
constants $c_i$'s are known, the estimator \eqref{eq:mainest}
resembles the balance heuristic estimator of \cite{veac:guib:1995},
which is discussed in \cite{owen:zhou:2000} as the deterministic
mixture. On the other hand, in several
applications of IS methods, $\bd$ in \eqref{eq:mainest} is unknown,
which is the case when $Q=\Pi$, that is, when samples from a subset of
densities of $\Pi$ are used to estimate the normalizing constants for
the entire family via~\eqref{eq:mainest}. Routine applications of IS
estimation with $Q=\Pi$ can be found in Monte Carlo maximum likelihood
estimation, Bayesian sensitivity analysis and model selection
\citep{geye:thomp:1992, buta:doss:2011, doss:2010}. If $\bd$ is
unknown, \cite{doss:2010} proposed a two-stage method, where in the
first step, using samples from $q_i, i=1,\dots,k$, $\bd$ is
estimated by $\hatbd$ using \pcite{geye:1994} reverse logistic regression estimator
or \pcite{meng:wong:1996} bridge sampling method. Then, independent of
step one, new samples are used to calculate~\eqref{eq:mainest} with
$\bd$ replaced by $\hatbd$.

The effectiveness of \eqref{eq:mainest} depends on the choice of $k$,
$\ba$, $n_l$, and the importance densities $\bq =\{q_1, \dots, q_k\}$. This article focuses on the choice of the importance
densities because it is the most crucial, and the multiple IS
estimator \eqref{eq:mainest}, just like the naive IS estimator
\eqref{eq:stis}, is useless if these densities are `off
targets'. Although increasing $k$ or $n_l$, may lead to estimators
with less variance, it results in higher computational cost, therefore
these are often determined based on the available computational
resources. On the other hand, for fixed $k$, $\ba$, and $n_l$,
efficiency and stability of the estimator \eqref{eq:mainest} can be
highly improved by appropriately choosing the $k$ importance densities
$\bq$ from the set $Q$.

This paper is the first where systematic methods of selection of
proposal distributions for IS are developed and tested. We propose
three approaches. (i) Our first approach is based on a geometric
spatial design method, called the space filling (SF) method. In
particular, among all subsets $\bq \subset Q$ with $|\bq| =k$, the one
that minimizes the gaps between the elements of $\bq$ and those of
$\Pi$ is chosen. The choice of the distance between the elements of
$\bq$ and $\Pi$ is crucial, and here we propose the symmetric
Kullback-Leibler divergence.  (ii) The second approach, called the
minimax (MNX) method, chooses $\bq$ that minimizes the maximum
standard error, or the maximum relative standard error of the
estimator $\hat u$ (or $\hat{\eta}^{[f]}$). (iii) Finally, the third
approach is applicable when $\bd$ in \eqref{eq:mainest} is unknown,
and \pcite{doss:2010} two-stage IS method is used.  In this approach,
called the maximum entropy (ENT) method, following the maximum entropy
criterion of experimental design, $\bq$ is chosen by maximizing the
determinant of the asymptotic covariance matrix of $\hatbd$.  We
describe and compare these three methods in details in
Section~\ref{sec:sel}. Each of the three methods is better suited to
different situations. MNX is applicable to any IS estimator for which
valid standard errors are available. Implementation of both MNX and
ENT needs estimates of asymptotic variances in a central limit
theorem. In the absence of such variance estimates, SF can be used. SF
does not depend on the form of the particular IS estimator
\eqref{eq:mainest}, thus, the same SF proposal distributions are used
for any IS estimator. However, successful implementation of the SF, as
shown later, crucially depends on the choice of the metric. Unlike the
MNX design, which depends on the choice of the function $f$, the same
SF and ENT proposals work no matter if the goal is to estimate the
normalizing constants or the means. Overall, SF is the most
straightforward to implement, although SF may not always be ideal, as
it is independent of the form of the estimator and the particular
estimand of interest, in our experience, with a properly chosen
metric, it consistently provides desirable results.  The three methods
are implemented in the R package geoBayes \citep{R:geob}.  We
illustrate these methods using several detailed examples involving
autologistic models, Bayesian regression models and spatial
generalized linear mixed models.

Unfortunately, in the literature, there is not much discussion on the
choice of the importance densities in multiple IS methods, although
given $\bq$, in the special case when $\bd$ is known and iid samples
are available from the proposals, there are some methods for selecting
the weights $\ba$ \citep[see e.g.][]{li:tan:chen:2013}. One exception
is \cite{buta:doss:2011} who described an ad-hoc method in the
important special case of $Q=\Pi$. \cite{buta:doss:2011} stated that
solving the minimax variance design problem, that is, the one that
minimizes
$\phi(\bq) = {\mbox{max}}_{\pi \in \Pi} \sigma^2_u(\pi; \bq)$ exactly,
where $\sigma^2_u(\pi; \bq)$ is the asymptotic variance of $\hat u$
in~\eqref{eq:mainest}, is `hopeless'. Assuming that a consistent
estimator $\hat{\sigma}^2_u(\pi; \bq)$ of ${\sigma}^2_u(\pi; \bq)$ is
available, \cite{buta:doss:2011} proposed a procedure where starting
from some `trial' proposal pdfs, $\hat{\sigma}^2_u(\pi; \bq)$ is
computed for all $\pi \in \Pi$. Then, proposal densities are either
moved to regions of $\Pi$ where $\hat{\sigma}^2_u(\pi; \bq)$ is large,
or new proposal densities from these high variance regions are added
increasing $k$. Here, we develop a principled approach, called the
sequential method (SEQ), formalizing this procedure
and compare its
performance with the three proposed methods.

As mentioned above, the MNX and ENT approaches developed here as well
as the SEQ method utilize asymptotic standard errors of $\hatbd$ and
$\hat{u}$. Another contribution of this paper is the development of
spectral variance (SV) estimators of asymptotic variances for $\hatbd$
and $\hat{u}$.  Availability of consistent estimators is important in
its own right as it allows for calculation of asymptotically valid
standard errors of the IS estimators. Recently,
\cite{roy:tan:fleg:2018} provided standard errors estimators of
$\hatbd$ and $\hat{u}$ using the batch means method.  In different
numerical examples (not shown here), we observe that the proposed SV
estimators are generally less variable than the batch means
estimators. This observation is in line with \citet{fleg:jone:2010}
who showed that, for estimating means of scalar valued functions,
certain SV estimators are less variable than the batch means
estimators by a factor of 1.5.

The rest of the paper is organized as
follows. In Section~\ref{sec:imp}, we describe both the multiple IS
estimation as well as the reverse logistic regression estimation. The
proposed methods of selecting proposal densities for IS
estimators are described in Section~\ref{sec:sel}. Some illustrative
examples are given in Section~\ref{sec:exam}.
Section~\ref{sec:disc}
contains conclusions of the paper. Proofs of theorems and several examples are relegated to the supplementary
materials.

\section{Multiple IS estimation of normalizing constants and expectations}
\label{sec:imp}

\vspace{-.2in}

Recall that $\Pi = \{\pi: \pi(x) = \nu(x)/\theta\}$ is a family of
target densities on $\sX$, and $f: \sX \rightarrow \mathbb{R}$ is a
function of interest.  Given samples
$\Phi_l \equiv \{X_i^{(l)}\}_{i=1}^{n_l}, l=1\ldots,k$ from a small
number of proposal densities
$\{q_l = \varphi_l(x)/c_l, l = 1,\dots, k\}$, one wants to estimate
$\theta$ (or, rather $\theta/c_1$) and $E_\pi f$ for all
$\pi \in \Pi$. Recall that we estimate $u(\pi, q_1) \equiv \theta/c_1$
and $E_\pi f$ by $\hat{u}(\bd) \equiv \hat{u}(\pi; \bd)$ defined in
\eqref{eq:mainest} and
$\hat{\eta}^{[f]} \equiv \hat{\eta}^{[f]} (\pi; \bd)$,
respectively. We also consider the more general setting when $\bd$ is
unknown, which is the case if $Q=\Pi$.  In such situations, we use the
two-stage IS procedure of \cite{doss:2010}, where first, $\bd$ is
estimated using \pcite{geye:1994} reverse logistic regression 
method (described in Section~\ref{sec:revl}) based on Markov chain
samples $\tilde{\Phi}_l \equiv \{\tilde{X}_i^{(l)}\}_{i=1}^{N_l}$ with
stationary density $q_l$, for $l=1,\ldots,k$. Once $\hatbd$ is formed,
independent of stage 1, new samples
$\Phi_l \equiv \{X_i^{(l)}\}_{i=1}^{n_l}, l=1\ldots,k$ are obtained to
estimate $u(\pi, q_1)$ and $E_\pi f$ by $\hat{u}(\hatbd)$ and
$\hat{\eta}^{[f]} (\pi; \hatbd)$, respectively.
\citet{buta:doss:2011} quantify benefits of the two-stage scheme as opposed to using the
same samples  to estimate both $\bd$ and $u(\pi,
q_1)$.

\subsection{Reverse logistic regression estimator of $\bd$}
\label{sec:revl}

\vspace{-.2in}

Let
$N=\sum_{l=1}^k N_l$ and $a_l \in [0, 1]$ for $l = 1, \ldots, k$
such that $\sum_{l=1}^k a_l =1$. Define
\begin{equation}
\label{eq:defzeta}
  \zeta_l = -\log(c_l) + \log(a_l),\; l = 1, \ldots, k,
\end{equation}
and
\begin{equation}
  \label{eq:pl}
  p_l(x, \bzeta) = \frac{ \varphi_l(x) e^{\zeta_l} } { \sum_{s=1}^k
  \varphi_s(x) e^{\zeta_s} },\; l = 1, \ldots, k,
\end{equation}
where $\bzeta=(\zeta_1, \dots, \zeta_k)$. (Note that, if $a_l = N_l/N$,
given that $x$ belongs to the pooled sample
$\big\{ \tilde{X}_i^{(l)}, \, i = 1, \ldots, N_l, \, l = 1, \ldots, k \big\}$,
$p_l(x, \bzeta)$ is the probability that $x$ comes from the
$l^{\text{th}}$ distribution.) Following \cite{doss:tan:2014},
consider the log quasi-likelihood function
\begin{equation}
  \label{eq:new-lql}
  \ell_N(\bzeta) = \sum_{l=1}^k a_l \frac{N}{N_l} \sum_{i=1}^{N_l} \log \bigl(
  p_l(\tilde{X}_i^{(l)}, \bzeta) \bigr).
\end{equation}
Note that adding the same
constant to all $\zeta_l$'s leaves~\eqref{eq:new-lql} invariant. Let
$\bzeta^0 \in \Real^k$ denote the true $\bzeta$ normalized to add to
zero, that is,
$\bzeta^0_l = \bzeta_l - \bigl( \sum_{j=1}^k \bzeta_j
\bigr)/k$. Here, $\bzeta_l$ denotes the $l$th element of $\bzeta$. Note that the function
$g \colon {\Real}^{k} \rightarrow {\Real}^{k-1}$ that maps $\bzeta^0$
into $\bd$ is given by $  g(\bzeta) =( e^{\zeta_1 - \zeta_2} a_2/a_1, e^{\zeta_1 - \zeta_3} a_3/a_1, \linebreak \dots, e^{\zeta_1 - \zeta_k} a_k/a_1)^{\top}.$
We estimate $\bzeta^0$ by $\hatbzeta$, where
\[
  \hatbzeta = \argmax \ell_N(\bzeta) \;\mbox{subject to}\;\sum_{j=1}^k \zeta_j = 0,
  \]
and thus, obtain $\hatbd = g(\hatbzeta)$.

\section{Selection of proposal distributions}
\label{sec:sel}

\vspace{-.2in}

In this section we propose three criteria for selecting the proposal
distributions $\bq=\{q_1,\ldots,q_k\} \subset Q$ for efficient use of the
multiple IS estimators.
For $\bq \subset Q$, the proposed criterion is generally denoted by
$\phi(\bq)$ and the optimal set is obtained by:
\begin{center}
  Minimize $\phi(\bq)$ over $\bq \subset Q$.
\end{center}
We consider the case where the set $Q$ corresponds to a family of
densities parameterized by $\xi \in \Xi$, thus searching over $Q$ is
equivalent to searching over $\Xi$.  The variable $\xi$ can be
multidimensional and the range of $\xi$, in every direction, can be
infinite. Thus, for computational purposes, it may be required to
narrow down the potential region of search, depending on the
application. \cite{evan:roy:2019} considered the problem of
maximizing~\eqref{eq:mainest} with respect to $\xi$, which, as
mentioned in the Introduction, is the situation in empirical Bayes
methods, so they used Laplace approximations to identify the region
where the maximizer may lie. Thus, using Laplace approximations, as in
\cite{evan:roy:2019}, we can narrow $\Xi$ down to a search set
$\tilde{\Xi}$. In Section S10 of the
supplement, 
we demonstrate an alternative approach to choosing $\tilde{\Xi}$ using preliminary samples.

Solving the minimization problem is a research problem in its own
right. We implemented two algorithms for searching over $\tilde{\Xi}$,
the point-swapping algorithm of \cite{royl:nych:1998}, and a simulated
annealing algorithm. Details about these algorithms are given in
Section~S7 of the supplementary materials. The
point-swapping algorithm generally requires more iterations, so it is
more suited to cases where the design criterion $\phi$ can be computed
quickly after a swap, as is often the case for the SF method.

\subsection{Space filling approach}
\label{sec:spfill}

\vspace{-.2in}

In this method, among all subsets $\bq =\{q_1,\dots,q_k\}$ of $Q$, the
one that minimizes the gaps between the elements of $\bq$ and the elements of $\Pi$ is
chosen. For $\pi \in \Pi$, $q \in Q$, let $\Upsilon(\pi, q)$ be a
suitably chosen metric. Define
\[
\psi_p(\bq, \pi) = \Big( \sum_{q \in \bq} \Upsilon(\pi, q)^p\Big)^{1/p},
\]
as a measure of `closeness' of $\bq$ to $\pi$. Note that, for $ p<0$,
$\psi_p(\bq, \pi) \rightarrow 0$ if $\pi$ is let to converge to a point in
$\bq$. The design criterion is to choose $\bq$ to minimize
\[
\phi_\mathrm{SF}(\bq) = \Psi_{p,\tilde{p}} (\bq) = \Big( \sum_{\pi \in \Pi} \psi_p(\bq, \pi)^{\tilde{p}}\Big)^{1/\tilde{p}}
\]
over all subsets $\bq$ with $|\bq|=k$. In the limit
($p \rightarrow -\infty, \tilde{p} \rightarrow \infty$),
$\Psi_{p,\tilde{p}}$ is related to the minimax design. However, as
\cite{royl:nych:1998} illustrate, keeping $p$ and $\tilde{p}$ finite
allows us to quickly evaluate $\phi$ after a swap of the point-swapping algorithm. We use
$p=-30$, $\tilde{p}=30$ in our examples, which allows us to
obtain a near-minimax SF design. 

The choice of the metric $\Upsilon(\pi, q)$ is crucial. For instance,
in the binomial robit model with degrees of freedom parameter $\xi$
(see the example in Section S8 of the supplemental materials), the
family of target densities
$\Pi \equiv \{\pi_\xi(x) = \nu_\xi(x)/\theta_\xi: \xi \in \Xi\}$ is
indexed by the Student's $t$ degrees of freedom parameter $\xi$.
Here, the relevant geometry (with respect to $\xi$) in $\mathbb{R}$ is
not Euclidean. Indeed, degrees of freedom $\xi = 10^2$ and $10^3$ are
close, but $\xi = 0.5$ and $\xi = 1$ are not. Thus, the SF based on
the Euclidean distance metric (SFE) may not be appropriate unless the
indexing variable is a location parameter. The Euclidean distance is
also sensitive to reparameterizations of the family of proposal
distributions. Another choice is the information metric
\citep{kass:1989, rao:1982} which measures the distance between two
parametric distributions using asymptotic standard deviation units of
the best estimator. The Kullback-Leibler divergence generates the
information number through the information metric
\citep{ghos:dela:sama:2007}. In practice, it may be difficult to
implement the information metric although it seems to be appropriate
for the context. Here, we use the symmetric Kullback-Leibler
divergence (SKLD) although it is not a metric, and denote the
corresponding method by SFS. Thus,
\begin{equation}
  \label{eq:skld-a}
    \Upsilon(\pi, q) = \int_{\sX} \pi(x) \log \frac{\nu(x)}{\varphi(x)} \mu(dx) -  \int_{\sX} q(x) \log \frac{\nu(x)}{\varphi(x)} \mu(dx).
  \end{equation}
  In the special case when
  $\Pi \equiv \{\pi_\xi(x) = \nu_\xi(x)/c_\xi: \xi \in \Xi\}$, that
  is, the target family is indexed by some variable $\xi$, and
  $Q= \Pi$, the SKLD between $\pi_{\xi_1}(x)$ and $\pi_{\xi_2}(x)$, is
  \begin{subequations}
  \begin{alignat}{2}
    \Upsilon(\xi_1, \xi_2) 
 &= \int_{\sX} \pi_{\xi_1}(x) \log \frac{\nu_{\xi_1}(x)}{\nu_{\xi_2}(x)} \mu(dx) -  \int_{\sX} \pi_{\xi_2}(x) \log \frac{\nu_{\xi_1}(x)}{\nu_{\xi_2}(x)} \mu(dx) \label{eq:skld-b}\\
& = \frac{ \int_{\sX} \nu_{\xi_1}(x) \log \frac{\nu_{\xi_1}(x)}{\nu_{\xi_2}(x)} \mu(dx)}{\int_{\sX} \nu_{\xi_1}(x) \mu(dx)} -  \frac{\int_{\sX} \nu_{\xi_2}(x) \log \frac{\nu_{\xi_1}(x)}{\nu_{\xi_2}(x)}\mu(dx)}{\int_{\sX} \nu_{\xi_2}(x) \mu(dx)} \label{eq:skld-c}.
  \end{alignat}
\end{subequations}
The SKLD \eqref{eq:skld-a} is generally not available in closed form. We
use a modified Laplace method \citep{evan:zhu:smith:2011} to
approximate \eqref{eq:skld-c}, and we describe the method in Section S1. The second order approximation described in the supplement is
exact when $\pi_{\xi_1}$ and $\pi_{\xi_2}$ are any two Gaussian densities. If $\sX$ is discrete, or the target
distributions are far from Gaussian, a Monte Carlo estimate of
\eqref{eq:skld-b} can be used with samples from $\pi_{\xi_1}$ and
$\pi_{\xi_2}$. Indeed, for some examples considered here, we use the Monte
Carlo estimate of \eqref{eq:skld-b} to implement SFS.

 The SF method does not involve the form of any
particular IS estimator. When $Q=\Pi$, the uniform (with respect to the chosen metric)
selection of the proposal distributions attempts to guarantee that
each target density is close to at least one proposal
distribution. Also, the SF method is attractive, as generally an IS
estimator is used to simultaneously estimate several quantities of
interest, resulting in different optimal design criteria.

\subsection{Minimax approach}
\label{sec:seq}

\vspace{-.2in}

Our second method is the minimax (MNX) design based on minimizing the maximum
SE or relative SE of $\hat{u}(\pi, \hatbd)$ or
$\hat{\eta}^{[f]} (\pi; \hatbd)$ over $\pi \in \Pi$.  Consistency and
asymptotic normality of $\hatbd$, $\hat{u}(\pi; \hatbd)$ and
$\hat{\eta}^{[f]} (\pi; \hatbd)$ are described in Theorems~1,~2
and~3, respectively of \cite{roy:tan:fleg:2018}. Let $\sigma^2_u(\pi,\bq)$
denote the asymptotic variance of $\hat{u}(\pi, \hatbd)$
when the set of proposal densities is $\bq \subset Q$. Then, the
standard error is $\sigma_u(\pi,\bq)/\sqrt{n}$, where $n=\sum_{l=1}^k n_l$. The
minimax approach chooses $\bq$ to minimize the largest standard error or
the relative standard error, given, respectively by
\begin{equation*}
  \phi_\textrm{MNX}(\bq) = \max_{\pi \in \Pi} \sigma_u(\pi,\bq)/\sqrt{n},\
  \text{and}\
  \phi_\textrm{MNX}(\bq) = \max_{\pi \in \Pi}
  \sigma_u(\pi,\bq)/\{\sqrt{n}\hat{u}(\pi, \hatbd)\} .
\end{equation*}
Similar measures can be derived in the case of
$\hat{\eta}^{[f]} (\pi; \hatbd)$ with variance
$\sigma^2_\eta(\pi,\bq)$.  In the following, we discuss estimation of
the asymptotic variances $\sigma^2_u(\pi,\bq)$ and
$\sigma^2_\eta(\pi,\bq)$ of these estimators. Note that the ratios of
the normalizing constants ($\theta/c_1$) can take large values as
$\pi$ varies in $\Pi$, especially when $\sX$ is multi-dimensional. The
standard errors corresponding to the distributions with large ratios
tend to be larger, whereas these standard errors for the distributions
with small (relative) normalizing constants can potentially be large
relative to the value of the estimates. Thus, if the goal is to
estimate the parameters corresponding to largest normalizing constants
(as in the empirical Bayes methods, see e.g. \cite{roy:evan:zhu:2016}),
then the first criterion can be used, on the other hand, if one wants
to estimate $\theta$ for all $\pi \in \Pi$, then the second criterion
(relative standard error) may be preferred.


{\bf Spectral variance estimation in reverse logistic regression and
  multiple IS methods:} First, we provide an SV estimator of the
asymptotic covariance matrix of $\hatbd$, as it is needed for the
asymptotic variances of $\hat{u}(\pi; \hatbd)$ and
$\hat{\eta}^{[f]} (\pi; \hatbd)$. Also, SV estimator of Var$(\hatbd)$
is important in its own right, and is used in Section \ref{sec:entr}
in our third approach to selection of proposal distributions.

As in \cite{roy:tan:fleg:2018}, we assume that the Markov chains
$\Phi_l$, $\tilde{\Phi}_l$ are {\it polynomially ergodic} for
$l=1,\dots, k$. (The definition of polynomial ergodicity of Markov chains
can be found in \cite{roy:tan:fleg:2018}.)
They showed that if the Markov chain $\tilde{\Phi}_l$ is
polynomially ergodic of order $t>1$ for $l=1,\dots, k$, then $\hatbzeta$
and $\hatbd$ defined in section~\ref{sec:revl} are consistent and asymptotically normal as $N_1,\ldots,N_k
\rightarrow \infty$, that is, there exist matrices $B, \Omega \in
\Real^{k,k}$ and $D \in \Real^{k,k-1}$ such that
\[
  \sqrt{N} (\hatbzeta - \bzeta) \cd {\cal N} (0, U) \quad\mbox{and}\quad \sqrt{N} (\hatbd - \bd) \cd {\cal N}
  (0, V) ,
\]
where $U=B^{\dagger} \Omega B^{\dagger}$ and $V = D^{\top} U D$. Here, for a square
matrix $C$, $C^{\dagger}$ denotes its Moore-Penrose inverse. The matrices
$B$, $\Omega$ and $D$ are as defined in (2.7), (2.8), and (2.5) respectively in
\cite{roy:tan:fleg:2018}. Theorem~\ref{thm:CLT} below provides consistent SV estimators
of the asymptotic variances of $\hatbzeta$ and $\hatbd$.

We now introduce some notations. Assume $N_l\rightarrow \infty$ such
that $\lim N_l/N \in (0, 1)$ for $l=1,\dots, k$.  Recall that
$\hatbd = g(\hatbzeta)$, and its gradient at $\hatbzeta$ (in terms of
$\hatbd$) is
\begin{equation}
  \label{eq:D}
  \widehat D =
  \begin{pmatrix}
    \hat d_2    & \hat d_3    & \ldots & \hat d_k    \\
    -\hat d_2   & 0      & \ldots & 0      \\
    0      & -\hat d_3   & \ldots & 0      \\
    \vdots & \vdots & \ddots & \vdots \\
    0      & 0      & \ldots & -\hat d_k
  \end{pmatrix}.
\end{equation}

As in \cite{roy:tan:fleg:2018},
the $k \times k$ matrix $\widehat{B}$ is defined by
\begin{equation}
  \label{eq:Bhat}
  \begin{split}
    \widehat{B}_{rr} & = \sum_{l=1}^k a_l \biggl( \frac{1}{N_l}
                         \sum_{i=1}^{N_l} p_r(\tilde{X}_i^{(l)}, \hatbzeta)
                         \bigl[ 1 - p_r(\tilde{X}_i^{(l)}, \hatbzeta) \bigr]
                         \biggr) \text{ and}\\
    \widehat{B}_{rs} & = - \sum_{l=1}^k a_l \biggl( \frac{1}{N_l}
                         \sum_{i=1}^{N_l} p_r(\tilde{X}_i^{(l)}, \hatbzeta)
                         p_s(\tilde{X}_i^{(l)}, \hatbzeta) \biggr) \text{ for } r \neq s,
  \end{split}
\end{equation}
that is, $\widehat{B}$ denotes the matrix of second derivatives of
$- \ell_N(\bzeta)/N$ evaluated at $\hatbzeta$, where $\ell_N(\bzeta)$ is defined in \eqref{eq:new-lql}.
Set $Z^{(l)}_{i} = \left( p_1(\tilde{X}_i^{(l)}, \hatbzeta), \ldots,
  p_k(\tilde{X}_i^{(l)}, \hatbzeta) \right) ^{\top}$ for $i=1,\dots, N_l$ and
$\bar{Z}^{(l)} =\sum_{i=1}^{N_l} Z^{(l)}_{i}/N_l$. Define the lag $j$ sample autocovariance as
\begin{equation}
\label{eq:gam}
\gamma_N^{(l)}(j) = \frac{1}{N_l} \sum_{i \in S_{j,N}} \left[ Z^{(l)}_{i} - \bar{Z}^{(l)} \right] \left[ Z^{(l)}_{i+j} - \bar{Z}^{(l)} \right]^{\top} \; \mbox{ for} \;\; l =1,\ldots, k,
\end{equation}
where 
$S_{j,N} = \{1, \dots, N-j\}$ for $j \ge 0$ and $S_{j,N} = \{(1-j),\dots,N\}$ for $j < 0$.
Let
\begin{equation}
\label{eq:SV}
\widehat{\Sigma}^{(l)} =  \sum_{j = -(b_{N_l} -1)}^{b_{N_l} -1} w_{N_l}(j) \gamma_N^{(l)}(j),
\end{equation}
where $w_{N_l} (\cdot)$ is the lag window, $b_{N_l}$'s are the
truncation points for $l =1,\ldots, k$.
 Finally, define
\begin{equation}
  \label{eq:Omegahat}
\widehat{\Omega} = \sum_{l=1}^k \frac{N}{N_l} a^2_l \widehat{\Sigma}^{(l)}.
\end{equation}

\begin{theorem}
  \label{thm:CLT}
  Assume that the Markov chains $\tilde{\Phi}_1, \ldots, \tilde{\Phi}_k$ are
  polynomially ergodic of order $t> 1$, and for all $l= 1,\ldots,k$,
  $w_{N_l}$ and $b_{N_l}$ satisfy conditions 1-4 in \citet[][Theorem
  2]{vats:fleg:jone:2015}. 
  Let $\widehat{D}$, $\widehat{B}$ and $\widehat{\Omega}$ be the matrices
  defined by~\eqref{eq:D}, \eqref{eq:Bhat}
  and~\eqref{eq:Omegahat}, respectively.  Then, as $N_l\rightarrow \infty$ for all $l= 1,\ldots,k$,
  $\widehat{U} := \widehat{B}^{\dagger}
  \widehat{\Omega} \widehat{B}^{\dagger}$ and
  $\widehat{V} := \widehat{D}^{\top} \widehat{U} \widehat{D}$ converge almost surely to
  $U$ and $V$, respectively.
\end{theorem}


Next, we consider estimation of the asymptotic variances of
$\hat{u}(\pi; \hatbd)$ and $\hat{\eta}^{[f]} (\pi;
\hatbd)$. \cite{roy:tan:fleg:2018} showed that, under certain
conditions, there exist $\sigma^2_u, \sigma^2_\eta > 0$ such that, as
$n_1,\ldots,n_k \rightarrow \infty$,
\begin{equation}
  \label{eq:cltueta}
  \sqrt{n} (\hat{u}(\pi; \hatbd) - u(\pi, q_1)) \cd N(0,
\sigma^2_u) \quad\mbox{and}\quad \sqrt{n} (\hat{\eta}^{[f]}(\pi;  \hatbd) -
E_\pi f) \cd N(0, \sigma^2_{\eta}).
\end{equation}
In Theorem~\ref{thm:elnc:elex} we provide consistent SV estimators of $\sigma^2_u$ and $\sigma^2_\eta$.
We first introduce some notations. Let
\begin{equation}
  \label{eq:uv}
 u^{\pi}(x; \bd) := \frac{\nu(x)} {\sum_{s=1}^k a_s \varphi_s(x) / d_s}
  \quad \text{and} \quad v^{[f], \pi}(x; \bd) := f(x) u^{\pi}(x; \bd).
\end{equation}
Define the vectors $c(\pi;\bd)$ and $e(\pi;\bd)$ of length $k-1$ with
$(j-1)$th coordinate as
\begin{gather}
  \label{eq:cdef}
   [c(\pi;  \bd)]_{j-1} = \frac{u(\pi, q_1)}{ d_j^2} \int_{\sX} \frac{a_j
     \varphi_{j} (x)}{\sum_{s = 1}^k a_s \varphi_{s} (x)/d_s} \pi (x) \mu(dx) \\
   \label{eq:defe}
 [e(\pi;  \bd)]_{j-1} = \frac{a_j}{d_j^2} \int_{\sX} \frac{[f(x) - E_\pi f]
  \varphi_{j} (x)}{\sum_{s = 1}^k a_s \varphi_{s} (x)/d_s} \pi(x) \mu(dx) ,
\end{gather}
for $j=2,\dots,k$, and their estimators
$\hat{c}(\pi;  \bd)$ and  $\hat{e}(\pi;  \bd)$ as
\begin{gather}
  \label{eq:chatdef}
   [\hat{c}(\pi;  \bd)]_{j-1} = \sum_{l=1}^k \frac{1}{n_l}
\sum_{i=1}^{n_l} \frac{a_j a_l \nu (X_i^{(l)}) \varphi_{j}
  (X_i^{(l)})}{ (\sum_{s = 1}^k a_s \varphi_{s} (X_i^{(l)})/d_s)^2 d_j^2} , \\
  \label{eq:defhate}
  [\hat{e}(\pi;  \bd)]_{j-1}  = \frac{\sum_{l=1}^k \frac{a_l}{n_l}
    \sum_{i=1}^{n_l} \frac{a_j f(X_i^{(l)}) \nu
      (X_i^{(l)}) \varphi_{j} (X_i^{(l)})}{d_j^2 (\sum_{s =
        1}^k a_s \varphi_{s} (X_i^{(l)})/d_s)^2}}{\hat{u} (\pi;  \bd)} - \frac{[\hat{c}(\pi;  \bd)]_{j-1} \hat{\eta}^{[f]} (\pi;  \bd)}{\hat{u} (\pi;  \bd)}.
\end{gather}
Suppose $b_{n_l}$'s are the truncation points, $w_{n_l}(j)$'s are lag
window, $u_{i} \equiv u_{i}(\bd) \equiv u^{\pi}(X_i^{(l)}; \bd)$,
$v^{[f]}_{i} \equiv v^{[f]}_{i}(\bd)\equiv v^{[f], \pi}(X_i^{(l)};
\bd)$, and $\bar{u} \equiv \bar{u}(\bd)$,
$\bar{v}^{[f]} \equiv \bar{v}^{[f]}(\bd)$ are the averages of
$\{u^{\pi}(X_{1}^{(l)}; \bd), \cdots, u^{\pi}(X_{n_l}^{(l)}; \bd)\}$
and
$\{v^{[f], \pi}(X_{1}^{(l)}; \bd), \cdots, v^{[f], \pi}(X_{n_l}^{(l)};
\bd)\}$, respectively. (Note that, abusing notations, the dependence
on $l$ is ignored in $u_{i}, v^{[f]}_{i}, \bar{u}$ and
$\bar{v}^{[f]}$.) Let
\begin{equation}
  \label{eq:tauldef}
  \hat{\tau}^2_l(\pi ;  \bd)= \frac{1}{n_l} \sum_{j = -(b_{n_l}
    -1)}^{b_{n_l} -1} w_{n_l}(j) \sum_{i \in S_{j,n}} \left[u_{i} -
    \bar{u}\right] \left[ u_{i+j} - \bar{u} \right], \;\mbox{and}
\end{equation}
\[
  \widehat{\Gamma}_{l} (\pi ;  \bd) = \frac{1}{n_l} \sum_{j = -(b_{n_l} -1)}^{b_{n_l} -1} w_{n_l}(j) \sum_{i \in S_{j,n}} \Bigg[ \left(
  \begin{array}{c}
    v_{i}^{[f]}\\
    u_{i}\\
  \end{array}
\right) -  \left(
  \begin{array}{c}
    \bar{v}^{[f]}\\
    \bar{u}\\
  \end{array}
\right) \Bigg] \Bigg[ \left(
  \begin{array}{c}
    v_{i+j}^{[f]}\\
    u_{i+j}\\
  \end{array}
\right) -  \left(
  \begin{array}{c}
    \bar{v}^{[f]}\\
    \bar{u}\\
  \end{array}
  \right) \Bigg]^{\top}.
  \]
  Finally, let
  $\hat{\tau}^2 (\pi ; \bd) = \sum_{l =1}^k (a_l^2 n/n_l)
  \hat{\tau}^2_l(\pi ; \bd)$,
  $\widehat{\Gamma} (\pi ; \bd) = \sum_{l =1}^k (a_l^2n/n_l)
  \widehat{\Gamma}_l(\pi ; \bd)$, and
  \[\hat{\rho}(\pi; \hatbd) = \nabla h(\hat{v}^{[f]}(\pi ; \hatbd),
  \hat{u}(\hatbd))^{\top} \widehat{\Gamma}(\pi; \hatbd) \nabla
  h(\hat{v}^{[f]}(\pi ; \hatbd), \hat{u}(\hatbd)),\] where
  $\nabla h(x, y) = (1/y, -x/y^2)^{\top}$.

\begin{theorem}
  \label{thm:elnc:elex}
  Suppose that for $\tilde{\Phi}_l, l=1,\dots,k$, conditions of
  Theorem~\ref{thm:CLT} hold and $\widehat{V}$ is the consistent SV
  estimator of $V$. Suppose that $N_l, n_l \rightarrow \infty$ for all $l=
  1,\ldots,k$, and there exists $\varpi \in [0, \infty)$ such that
  $n/N \rightarrow \varpi$. 
   In
  addition, let $n_l/n \rightarrow s_l \in (0, 1)$ for $l = 1,\cdots,k$.
Assume that the Markov chains
  $\Phi_1, \ldots, \Phi_k$ are polynomially ergodic of order
  $t \ge (1+\epsilon)(1 + 2/\delta)$ for some $\epsilon, \delta >0$
  such that
  $E_{q_l} |u^{\pi}(X;  \bd)|^{4+\delta} < \infty$, and for each
  $l = 1,\cdots,k$, $w_{n_l}$ and $b_{n_l}$ satisfy conditions 1-4 in \citet[][Theorem
  2]{vats:fleg:jone:2015}.
  \begin{enumerate}
  \item[(a)] Then $\hat{\sigma}^2_u = (n/N) \hat{c}(\pi; \hatbd)^{\top} \widehat{V} \hat{c}(\pi; \hatbd)
    + \hat{\tau}^2 (\pi ; \hatbd)$ converges almost surely to 
    $\sigma^2_u$.
  \item[(b)] In addition, suppose that
    $E_{q_l} |v^{[f], \pi}(X; \bd)|^{4+ \delta} < \infty$. Then
    $\hat{\sigma}^2_\eta = (n/N) \hat{e}(\pi; \hatbd)^{\top} \widehat{V} \hat{e}(\pi; \hatbd) +
    \hat{\rho} (\pi ; \hatbd)$ 
    converges almost surely to $\sigma^2_{\eta}$.
  \end{enumerate}
\end{theorem}
The estimators $\widehat{V}$ as well as $\hat{\sigma}^2_u$ and
$\hat{\sigma}^2_\eta$ are implemented in the R package geoBayes
\citep{R:geob}.  Since samples are obtained by running the Markov
chains with the stationary densities in $\bq$, we denote the
corresponding reverse logistic regression estimator of
$\bd \equiv \bd_{\bq}$ by $\hatbd_{\bq}$ and its asymptotic variance
as ${V}_{\bq}$. Similarly, in this case, we denote the SV estimators
of the asymptotic variances \eqref{eq:cltueta} of
$\hat{u}(\pi; \hatbd_{\bq})$ and $\hat{\eta}^{[f]}(\pi; \hatbd_{\bq})$
as $\hat{\sigma}^2_u (\pi; \bq )$ and
$\hat{\sigma}^2_\eta (\pi; \bq)$, respectively.

When $Q = \Pi$, a less computationally demanding approach is the SEQ
method in which densities are chosen sequentially from $\Pi$ where
$\hat{\sigma}^2_u (\pi; \bq )$ is the largest. Specifically, starting
with an initial density $\bq_1 = \{\tilde{q}\}$, suppose that we have
completed the $i$th step with the set $\bq_i$ chosen along with
(Markov chain) samples from each density in $\bq_i$. If $\bd$ is
unknown, part of this sample (stage 1) is used for calculating the 
estimator $\hatbd$, and the remaining sample is used to compute
$\hat\sigma^2_u(\pi; \bq_i)$ for the remaining densities
$\pi \in \Pi \setminus \bq_{i}$. Then
$\bq_{i+1} = \bq_{i} \cup \{\pi_j\}$ where
$\pi_j = \argmax_{\pi \in \Pi \setminus \bq_i} \hat\sigma^2_u(\pi;
\bq_i)$, and the existing (Markov chain) sample is augmented with
samples from $\pi_j$. Thus, at each step, the density corresponding to
the largest (estimated) asymptotic variance is chosen. The process is
repeated until $k$ densities have been selected.  The initial
$\tilde{q}$ can be the density where the multiple IS estimator
\eqref{eq:mainest} or any other interesting quantity based on samples
from a preliminary SF set is maximized (see Section S10 of the
supplement 
for an example). 

\subsection{Maximum entropy approach}
\label{sec:entr}

\vspace{-.2in}

The third method uses maximum entropy sampling \citep{shew:wynn:1987}
for selecting $\bq$.  This method is applicable when $\bd$ is unknown and is
developed in the context of \pcite{doss:2010} two-stage IS estimation
scheme described before. We use the notation $\mbox{Ent}(\cdot)$ to
denote the Boltzmann-Shannon entropy of the random variable inside the
brackets.
The %
maximum entropy (ENT) approach chooses $\bq$ that minimizes
\begin{equation*}
  \phi_\mathrm{ENT}(\bq) = -\mbox{Ent}(\hatbd_{\bq}).
\end{equation*}
This is interpreted as sampling those elements of $Q$ that carry the most
uncertainty in $\hatbd_{\bq}$. As we show below, since $\hatbd_{\bq}$ is used in the
calculation of both $\hat{u}$ and $\hat{\eta}^{[f]}$, the optimal $\bq$
will cause (asymptotically) lower uncertainty in those estimators.
Note that since $\hatbd_{\bq}$ depends on
the reference density $q_{1}$, it is assumed that $q_1$ remains
fixed, which can be the density $\tilde{q}$ discussed in Section~\ref{sec:seq}.
In the following, we assume that the objective
is to estimate ratios of normalizing constants. In the supplementary
materials, we derive similar results under the objective of estimating means
$E_\pi f$.

To derive a formula for $\mbox{Ent}(\hatbd_{\bq})$ we require the
asymptotic joint distribution of $\hatbd_{\bq}$ with $\hat u$ over $\Pi$.
Let $\hat{\bf u}(\bpi; \hatbd_{\bq})$ be the vector of length $|\Pi|$
consisting of $\hat{u}(\pi; \hatbd_{\bq})$'s, $\pi \in \Pi$ in a (any)
fixed order. Indeed, we refer to this fixed ordering whenever we write
$\Pi$ in this section. Similarly define the vector of true (ratios of)
normalizing constants ${\bf u}(\bpi, q_1)$. Let $C(\bpi; \bd_{\bq})$ be the
$|\Pi| \times (k-1)$ matrix with rows $c(\pi; \bd_{\bq})$ (defined
in~\eqref{eq:cdef}), $\pi \in \Pi$. Similarly, define
$\widehat{C}(\bpi; \bd_{\bq})$ with rows $\hat{c}(\pi; \bd_{\bq})$ (defined
in~\eqref{eq:chatdef}), $\pi \in \Pi$. Let ${\bf u}^{\bpi}(x ;\bd_{\bq})$
be the $|\Pi|$ dimensional vector consisting of $u^\pi(x; \bd_{\bq})$'s
defined in \eqref{eq:uv}. Let $T_l(\bd_{\bq})$ be the $|\Pi| \times |\Pi|$
matrix with elements
\begin{align}
  \label{eq:tauxil}
  \tau^2_l(\pi, \pi' ; \bd_{\bq}) &= \Cov_{q_l} (u^\pi(X_1^{(l)}; \bd_{\bq}), u^{\pi'}(X_1^{(l)}; \bd_{\bq}))\\ & + \sum_{g=1}^{\infty}
\Cov_{q_l} (u^\pi(X_1^{(l)};  \bd_{\bq}), u^{\pi'}(X_{1+g}^{(l)};  \bd_{\bq})) + \sum_{g=1}^{\infty}
\Cov_{q_l} (u^\pi(X_{1+g}^{(l)};  \bd), u^{\pi'}(X_{1}^{(l)};  \bd_{\bq})). \nonumber
\end{align}
Finally, let
\begin{equation}
  \label{eq:tauxildef}
  \widehat{T}_l(\bd_{\bq})= \frac{1}{n_l}   \hspace{-.05in} \sum_{j = -(b_{n_l} -1)}^{b_{n_l} -1} \hspace{-.19in} w_{n_l}(j) \sum_{i \in S_{j,n}} \left[{\bf u}^{\bpi}(X_{i}^{(l)}; \bd_{\bq}) - \bar{{\bf u}}(
    \bd_{\bq})\right] \left[{\bf u}^{\bpi}(X_{i+j}^{(l)}; \bd_{\bq}) - \bar{{\bf u}}(
    \bd_{\bq})\right]^{\top}, \hspace{-.05in}
\end{equation}
where $b_{n_l}$'s are the truncation points, $w_{n_l}(j)$'s are the lag
windows, and
$\bar{\bf u}(\bd_{\bq})= \sum_{i=1}^{n_l}{\bf u}^{\bpi}(X_{i}^{(l)};  \bd_{\bq})/n_l$.
\begin{theorem}
  \label{thm:jtclt}
   Suppose that $N_l, n_l \rightarrow \infty$ for all $l=
  1,\ldots,k$, and there exists $\varpi \in [0, \infty)$ such that
  $n/N \rightarrow \varpi$. 
  In addition, let $n_l/n \rightarrow s_l \in (0,1)$ for
  $l = 1,\cdots,k$.
\begin{enumerate}
\item[(a)] Assume that the stage 1 Markov chains $\tilde{\Phi}_l, l =1,\dots,k$ are polynomially ergodic of order $t
  >1$. Further, assume that the stage 2 Markov chains
  $\Phi_l, l=1,\dots,k$ are polynomially ergodic of order $t$, and
  for some $\delta >0$
  $E_{q_l} |u^\pi(X;\bd_{\bq})|^{2+\delta} < \infty$ for each
  $\pi \in \Pi$ and $l = 1,\cdots,k$ where
  $t > 1+2/\delta$. Then as $n_1, \ldots,n_k \rightarrow \infty$,
  \begin{equation}
    \label{eq:multclt}
  \sqrt{n} \left(\begin{array}{c}
    \hatbd_{\bq} - \bd_{\bq}\\
    \hat{\bf u}(\bpi; \hatbd_{\bq}) - {\bf u}(\bpi, q_1)\\
\end{array} \right)
\cd N\Bigg(0, \left(\begin{array}{cc}
    \varpi V_{\bq} & \Sigma_{12}\\
    \Sigma_{21} & \Sigma_{22}\\
\end{array} \right)\Bigg),
  \end{equation}
where $\Sigma_{21} = \varpi C(\bpi; \bd_{\bq}) V_{\bq}$,
$\Sigma_{12} = \Sigma_{21}^{\top}$, and $\Sigma_{22} = \varpi C(\bpi; \bd_{\bq}) V_{\bq} C(\bpi; \bd_{\bq})^{\top} + \sum_{l=1}^k (a_l^2/s_l)T_l(\bd_{\bq})$.
\item[(b)] Suppose that the conditions of Theorem~\ref{thm:CLT} hold
  for the stage 1 Markov chains. Let $\widehat{V}_{\bq}$ be the
  consistent estimator of $V_{\bq}$ given in Theorem~\ref{thm:CLT}.
  Assume that the Markov chains $\Phi_l, l=1,\dots, k$ are
  polynomially ergodic of order $t \ge (1+\epsilon)(1 + 2/\delta)$ for
  some $\epsilon, \delta >0$ such that
  $E_{q_l} \|{\bf u}^{\bpi}(X ;\bd_{\bq})\|^{4+\delta} < \infty$,
  ($\|\cdot\|$ denotes the Euclidean norm) for all $l= 1,\ldots,k$,
  and $w_{n_l}$ and $b_{n_l}$ satisfy conditions 1-4 in
  \citet[][Theorem 2]{vats:fleg:jone:2015}. Then
  $(n/N) \widehat{C}(\bpi; \hatbd_{\bq}) \widehat{V}_{\bq}
  \widehat{C}(\bpi; \hatbd_{\bq})^{\top} + \sum_{l=1}^k
  (a_l^2/s_l)\widehat{T}_l(\hatbd_{\bq})$ and
  $(n/N) \widehat{C}(\bpi; \hatbd_{\bq}) \widehat{V}_{\bq}$ converges
  almost surely to $\Sigma_{22}$ and $\Sigma_{21}$, respectively.
  \end{enumerate}
\end{theorem}

Let
$Y \equiv (Y^T_{\bq},Y^T_{\Pi})^T$ be a random vector having the normal distribution in
\eqref{eq:multclt}. The Boltzmann-Shannon entropy of $Y$
is $\mbox{Ent}(Y) = \mbox{constant} + \frac{1}{2} \log \mbox{det}(\Sigma)$,
where $\Sigma$ is the covariance matrix of $Y$.
Note that
 \begin{align*}
   \log \mbox{det}(\Sigma) 
=  \log \mbox{det}(\varpi V_{\bq}) + \log \mbox{det}(\Sigma_{22} - \varpi C(\bpi; \bd_{\bq}) V_{\bq} C(\bpi; \bd_{\bq})^{\top}),
\end{align*}
where the second matrix on the right side is the covariance matrix of
the conditional distribution of
$Y_{\Pi}| Y_{\bq}$. Since
Theorem~\ref{thm:jtclt}~(b) provides a consistent estimator of this
conditional covariance matrix, we can minimize the determinant of this
estimator matrix to choose $\bq$.

As mentioned in \cite{shew:wynn:1987}, great computational benefit can be
achieved by converting this conditional problem to an unconditional
problem. In particular, as noted in \cite{shew:wynn:1987}, minimization of
the second term is equivalent to maximization of
$\log \mbox{det}(V_{\bq})$. 
In practice, we
would replace $V_{\bq}$ by its estimator given in Theorem~\ref{thm:CLT},
i.e. $\widehat{V}_{\bq}$, 
using Markov chain samples from densities in $\bq$. In this case, the ENT
criterion simplifies to
\begin{equation*}
  \phi_\mathrm{ENT}(\bq) = -\log\det(\widehat{V}_{\bq}).
\end{equation*}

Unlike the SF, MNX and SEQ methods, the ENT approach is applicable
only in the context of \pcite{doss:2010} two-stage IS estimation
scheme. In contrast, if the multiple IS estimator \eqref{eq:mainest}
is used, since ENT avoids the second stage IS estimation, it needs
fewer samples than the MNX and SEQ methods which require enough
samples to be used for both stages.  Also, ENT avoids computing the
target un-normalized densities $\nu$ for $\pi \in \Pi$. However, one
advantage of the MNX and SEQ methods is that at the end of the
procedure, we already have available samples from densities in $\bq$
which can be used in the two-stage IS estimation scheme.

\section{Examples}
\label{sec:exam}

\vspace{-.2in}

{\bf Autologistic model:} Consider the popular
autologistic models \citep{besa:1974}, which are Markov
random field models for binary observations. Let $s_i$ denote the
$i$th spatial location, and let
$\mbox{nb}_i \equiv \{s_j: s_j \mbox{ is a neighbor of}\; s_i\}$
denote the neighborhood set of $s_i, i =1, \dots, m$. Markov random
field models for $\bx = \{x(s_i), i=1,\dots,m\}$ are formulated by
specifying the conditional probabilities
$p_{i} = P(x(s_i) =1 | \{x(s_j): j \neq i\}) = P(x(s_i) =1 | \{x(s_j):
s_j \in \mbox{nb}_i\}), i=1,\dots,m$. For simplicity, we impose that
all neighborhoods have the same size
$w = |\mbox{nb}_i|, i =1, \dots, m$. We consider a centered
parameterization \citep{kais:cara:2012} given by
$\mbox{logit}(p_{i}) = \mbox{logit}(\kappa)+ (\gamma/w) \sum_{s_j \in
  \mbox{nb}_i} (x(s_j) - \kappa)$, where
$\mbox{logit}(z) = \log(z/[1-z])$, $\gamma$ is a dependence
parameter, and $\kappa$ is the probability of observing one in
the absence of statistical dependence.  Jointly, the probability mass function (pmf)
$\pi(\bx | \gamma, \kappa)$ of $\bx$ is given by (see Section
S9.1 
of the supplementary materials)
\begin{equation}
  \label{eq:autolog}
  \pi(\bx | \gamma, \kappa) \propto \exp\big\{ (\mbox{logit}(\kappa) - \gamma \kappa) \sum\nolimits_{i=1}^m x(s_i)+ \frac{\gamma}{2w} \sum\nolimits_{i=1}^m \sum\nolimits_{s_j \in
    \mbox{nb}_i} x(s_i) x(s_j)\big\}.
\end{equation}
The normalizing constant $\theta \equiv \theta(\gamma, \kappa)$ in
$\pi(\bx | \gamma, \kappa)$ is intractable when $\gamma \neq
0$. \cite{sher:apan:carr:2006} mention that `there is no known simple
way to approximate this normalizing constant'. Here, we use multiple
IS for estimating $\theta$ and then estimate $\xi = (\gamma,\kappa)$
by maximum likelihood method.

We consider a 10 $\times$ 10 square lattice on a torus, with
four-nearest (east-west, north-south) neighborhood structure with the
family of autologistic pmfs
$\Pi = \{\pi(x|\gamma,\kappa): \gamma = -4, -3.2, \ldots, 4, \kappa =
0.1, 0.2, \ldots, 0.9\}$. The family of importance densities $Q = \Pi$
in this case, therefore, choosing the importance densities amounts to
choosing the parameters $\xi$. We want to choose $k=5$ densities from
$Q$, i.e, $k$ different $\xi$ values, one of which must be
$\xi_1 = (0,0.5)$. We apply the multiple IS based on proposal
densities from the five methods, namely, SFE, SFS, MNX, SEQ, and ENT,
as well as the naive IS method NIS. MNX and SEQ are based on the
relative standard error criterion. Computation of the SFS, MNX, SEQ,
and ENT criteria is based on 20,000 stage 1 and 20,000 stage 2 samples
produced from each candidate density via Gibbs sampling (except for
the case $\gamma=0$ where independent sampling was used), after a burn
in of 4,000 each time. For the computation of the SV estimator we used
the Tukey-Hanning window (see Section S9.2). We observe that SEQ
chooses a skeleton set on the boundary of the search space for
$(\gamma,\kappa)$, while SFS, MNX, and ENT choose some points close to
the boundary (see Section S9.2).

To test the performance of the different methods when used to estimate
the parameters $\xi$, we simulate from the model for different choices
of $\xi$ as shown in Table~\ref{tab:bias_rmse_autologistic}, and then
estimate these parameters using the maximum likelihood method. As the
likelihood is intractable, $\theta(\gamma,\kappa)/\theta(0,0.5)$ is
estimated via \eqref{eq:mainest} with the proposal densities derived
from each method. To that end, we took 10,000 samples from each
density after a burn in of 1,000. For NIS we took 50,000 samples. We
generated 125 realisations (data) for each choice of
$(\gamma, \kappa)$ parameters. We observed that some realized data
resulted in an unbounded likelihood for some methods. NIS was most
affected with 39\% of the realized values resulting in an unbounded
likelihood followed by SEQ with 11\% and ENT with
8\%. Table~\ref{tab:bias_rmse_autologistic} shows the
root mean squared error for estimating $\gamma$ excluding the
cases with unbounded likelihood for each method. The results show that
the multiple IS methods perform significantly better than NIS. Between
the multiple IS methods, we note that SEQ has in general worse
performance than MNX and SFE is worse than SFS. The root
mean squared error for estimating $\kappa$ does not show significant
differences across the multiple IS methods so it is not shown,
although we observed that NIS performed worse.  Further comparisons
and computational details are given in Section~S9.2 in the
supplementary materials. 
\begin{table}[H]
  \centering
  \begin{tabular}{rr|rrrrrr}
    \hline
    $\kappa$ & $\gamma$ & NIS &  SFE
&  SFS &  MNX &  SEQ &  ENT \\ \hline
0.2 &   --\,1 &  7.55 & 3.68 & 4.14 & 4.62 &  5.19 & 3.65 \\
0.2 &    1  & 10.91 & 3.63 & 1.67 & 1.67 &  1.74 & 1.69 \\
0.3 &   --\,2  &  8.75 & 1.38 & 1.61 & 1.60 &  9.42 & 1.37 \\
0.3 &    2 &    5.13 & 1.17 & 1.19 & 1.18 &  1.21 & 1.18 \\
0.4 &   --\,3  &  4.51 & 5.36 & 1.59 & 1.52 &  9.55 & 1.63 \\
0.4 &    3  &  3.76 & 1.11 & 1.12 & 1.11 &  1.18 & 1.11 \\
0.5 &   --\,4  & 10.69 & 5.65 & 1.20 & 1.15 & 10.13 & 3.61 \\
0.5 &    4 &    4.83 & 1.04 & 1.04 & 1.03 &  1.06 & 1.02 \\
0.6 &   --\,3  &  6.71 & 1.33 & 1.21 & 1.22 &  6.16 & 7.59 \\
0.6 &    3 &    3.65 & 1.12 & 1.12 & 1.12 &  1.22 & 1.12 \\
0.7 &   --\,2  &  9.62 & 1.62 & 1.93 & 1.79 &  1.80 & 5.97 \\
0.7 &    2 &  6.09 & 1.27 & 1.27 & 1.26 &  1.35 & 1.48 \\
0.8 &   --\,1 & 14.84 & 5.37 & 4.52 & 3.64 &  4.38 & 5.88 \\
0.8 &    1 & 11.88 & 2.14 & 1.96 & 1.94 &  2.06 & 2.40 \\
\hline
  \end{tabular}
  \caption{Root mean squared error for estimating $\gamma$ in the autologistic
    example.}
  \label{tab:bias_rmse_autologistic}
\end{table}

\noindent{\bf Bayesian negative binomial regression:} We consider a
Bayesian negative binomial regression model with response variable
$y_i$, $i=1,\ldots,21$, generated independently from the negative
binomial distribution with size parameter $\xi$ and mean for $y_i$,
$\mu_i = \exp(\beta_0 + \beta_1 \times w_i)$,
$w_i = -1 + 0.1\times(i-1)$. Here, $x = (\beta_0, \beta_1 )$ are
unknown parameters, assigned a bivariate normal prior with mean 0 and
covariance matrix $10 (W^\top W)^{-1}$, where $W$ denotes the design
matrix. As $\xi \rightarrow \infty$, the negative binomial
distribution converges to the Poisson distribution. 
Let the family of target densities $\Pi$ be the posterior densities
for $x$ for $\xi \in (0,\infty]$. Here, $\xi = \infty$ corresponds to
the the Poisson model. We wish to compute the logarithm of Bayes factor
$b_\xi = \log(\theta_\xi/\theta_\infty)$, where $\theta_\xi$ denotes
the unknown normalizing constant of the posterior
density. 
The Bayes factor can be used to decide between the models for given
data. We estimate $b_\xi$ by multiple IS using
\eqref{eq:mainest}, with the proposal densities chosen from $\Pi$,
i.e.  $Q=\Pi$, one of which must correspond to $\xi = \infty$ and two
more densities chosen from $\tilde{\Xi} = \{1,2,\ldots,40\}$, i.e.,
$k=3$. The choice of the proposal densities 
for MNX and SEQ are based on the relative standard error of the
multiple IS estimator of $\exp(b_\xi)$. For comparison, we also
consider the naive IS (NIS) method with proposal at $\xi = \infty$. 

We generate data from four models with $\xi = 0.5, 1, 2, \infty$ and
$(\beta_0,\beta_1) = (1,0.5)$, 400 times from each model. For each
data set we compute the skeleton set for the 5 criteria: SFE, SFS,
MNX, SEQ, and ENT. We used $N_l=n_l=3,600$ Monte-Carlo samples from
the $l$th proposal, after a burn in of 1,000, $l=1,2,3$, for computing
the spectral variance estimates, and the SKLD was also computed using
the same samples. The Monte-Carlo algorithm was implemented using the
R package rstan \citep{stan:2017}. After the skeleton set for each
method and data set is found, we generate additional 5,000 Monte Carlo
samples from each proposal, discard the first 1,000, and use the
remaining 4,000 to compute the estimator of $b_\xi$ for all
$\xi \in \tilde{\Xi}$ via~\eqref{eq:mainest}. For NIS we used 12,000
samples in total from the proposal density. Alternatively,
$\theta_\xi$ can be computed by numerical integration. For this, we
use the Gauss-Kronrod method as implemented in the R package pracma
\citep{R:pracma} 
with relative error set to $10^{-6}$, from where we can compute
$b_\xi$.  We treat the estimates obtained by numerical integration as
the golden standard and compare each IS estimate against it. As the
models are very similar for large values of $\xi$, our comparison
concentrates in the range $\xi =1, \dots, 10$. The average root mean
squared difference between the IS estimate of $b_\xi$ for each method
and the one obtained via numerical integration for the 400 simulations
and over $\xi =1, \dots, 10$ are given in
Table~\ref{tab:NBexampleError}.  The results show that generally MNX
and ENT have better performance than SEQ, both for estimating the
Bayes factor and the regression coefficient and that SFS is better
than SFE. NIS performs significantly worse than the multiple IS
methods.\begin{table}[H] \centering
  \begin{tabular}{|r|rrrr|}
    \cline{1-5}
    &   0.5 &     1 &     2 &$\infty$\\ \hline
NIS & 1214.640 & 716.045 & 383.153 & 129.079\\
SFE & 2.916 & 2.698 & 2.080 & 2.172\\
SFS & 2.337 & 2.343 & 1.712 & 1.850\\
MNX & 2.222 & 2.161 & 1.594 & 1.806\\
SEQ & 2.293 & 2.307 & 1.745 & 1.810\\
ENT & 2.266 & 2.140 & 1.626 & 1.774\\
\hline
  \end{tabular}
  \caption{Average root mean squared difference between the estimates
    obtained by IS and the values obtained via numerical integration for
    $b_\xi$. 
    The table shows the original values
    multiplied by 100.}
  \label{tab:NBexampleError}
\end{table}

\section{Discussions}
\label{sec:disc}

\vspace{-.2in}

We consider situations where one is simultaneously interested in large
number of target distributions, as in model selection and sensitivity
analysis examples. Multiple IS estimators are particularly useful in this
context, however, the choice of proposal distributions for these estimators
has not received much attention in the literature. We provide three
systematic techniques for addressing this issue. The
first method, based on a geometric space filling criterion, and the
second method, based on the minimax asymptotic standard error, can be used for
any multiple IS estimators. The third, maximum entropy method, is designed for
the two-stage multiple IS estimators of \cite{doss:2010}. We
compare the performance of these three methods in several examples.
Our results show that careful choice of the proposal densities, as produced
by our methods, results in more accurate estimates.

The proposed minimax and entropy methods use asymptotic standard
errors for the multiple IS and the reverse logistic regression
estimators, respectively. We construct consistent SV estimators for
these standard
errors. 
These estimators are important in their own
right as they are valuable for assessing the quality of the
multiple IS estimators and the reverse logistic regression estimator. 


\newpage
\vspace{.55cm}
 \centerline{\bf Supplementary Material}
\vspace{.55cm}
\setcounter{section}{0}
\setcounter{equation}{0}
\def\theequation{S\arabic{section}.\arabic{equation}}
\def\thesection{S\arabic{section}}

\fontsize{12}{14pt plus.8pt minus .6pt}\selectfont

  \section{A modified Laplace approximation for Kullback-Leibler
divergence}
\label{sec:laplace}
In this section, we describe a modified Laplace approximation for the
symmetric Kullback-Leibler divergence (SKLD) defined in the paper. Let
$\sX = \mathbb{R}^r$, for some $r \ge 1$, and $\mu$ be the Lebesgue
measure. Consider the SKLD between two densities
$\pi_{\xi_1}(x) = \nu_{\xi_1}(x)/c_{\xi_1}$ and
$\pi_{\xi_2}(x) = \nu_{\xi_2}(x)/c_{\xi_2}$, with the assumption
$\log \nu_{\xi_i} (x) = O(M)$ for some $M, i=1, 2$. Note that
\begin{equation}
  \label{eq:skl}
  \Upsilon (\xi_1, \xi_2) = \frac{ M \int_{\sX} J(x) \exp(G(x)) \mu(dx)}{\int_{\sX} \exp(G(x)) \mu(dx)} -  \frac{M\int_{\sX} J(x) \exp(H(x)) \mu(dx)}{\int_{\sX} \exp(H(x)) \mu(dx)},
\end{equation}
where $G(x) = \log \nu_{\xi_1}(x)$, $H(x) = \log \nu_{\xi_2}(x)$, and
$J(x) = (G(x) - H(x))/M$. We apply Laplace approximation on each
integral in \eqref{eq:skl} separately. Specifically, we expand the
integrals in the first term around
$\hat{x} = \argmax_{x \in \sX} G(x)$ and the integrals in the second
term around $\tilde{x} = \argmax_{x \in \sX} H(x)$. Let $\hat{J}$
and $\tilde{J}$ denote $J$ evaluated at $\hat{x}$ and $\tilde{x}$
respectively. We denote
$\hat{G}_i = \frac{\partial}{\partial x_i} G(x) |_{x= \hat{x}}$ and
similarly $\hat{G}_{ij}$ for second order partial derivatives and so
on. We also denote $\hat{G}^{-1}_{ij}$ to be the $(i,j)$th element of
the inverse of the matrix with elements $\hat{G}_{ij}$'s. Then by an
application of (17) from \cite{evan:zhu:smith:2011}, we have
\[
  \frac{\int_{\sX} J(x) \exp(G(x)) \mu(dx)}{\int_{\sX} \exp(G(x))
    \mu(dx)} \approx \hat{J} + \frac{1}{2} \hat{J}_{i_1} \hat{G}_{i_2
    i_3 i_4} \hat{G}^{-1}_{i_1 i_2} \hat{G}^{-1}_{i_3 i_4} - \frac{1}{2} \hat{J}_{i_1 i_2} \hat{G}^{-1}_{i_1 i_2}
\]
with an implicit summation $i_1,\dots, i_4 \in \{1, \dots, r\}$. A similar approximation
is derived for the second term:
\[
  \frac{\int_{\sX} J(x) \exp(H(x)) \mu(dx)}{\int_{\sX} \exp(H(x))
    \mu(dx)} \approx \tilde{J} + \frac{1}{2} \tilde{J}_{i_1}
  \tilde{H}_{i_2 i_3 i_4} \tilde{H}^{-1}_{i_1 i_2} \tilde{H}^{-1}_{i_3 i_4} -
  \frac{1}{2} \tilde{J}_{i_1 i_2} \tilde{H}^{-1}_{i_1 i_2}.
\]
The first order approximation to $\mbox{SKLD} (\xi_1, \xi_2)$ is
$M(\hat{J} - \tilde{J})$, which may be sufficient, but not if
$\hat{x} = \tilde{x}$. Note that, the second order approximation is
exact for two Gaussian densities. 

\section{Proof of Theorem 1}
\label{sec:appthm1}

From \cite{roy:tan:fleg:2018}, we only need to show $\widehat{\Omega} \cas
  \Omega$ where the SV estimator $\widehat{\Omega}$ is
  defined in~(3.12) 
 and the $k \times k$ matrix $\Omega$, following \cite{roy:tan:fleg:2018}, is defined through
\begin{equation*}
  \label{eq:Omega}
 \Omega_{rs} =
  \sum_{l=1}^k \frac{a_l^2}{\tilde{s}_l} \Big[  E_{q_l}\{Y_1^{(r,l)} Y_1^{(s,l)}\} +  \sum_{i=1}^{\infty} E_{q_l}\{Y_1^{(r,l)} Y_{1+i}^{(s,l)}\}+  \sum_{i=1}^{\infty} E_{q_l}\{Y_{1+i}^{(r,l)} Y_1^{(s,l)}\}\Big],
\end{equation*}
for $r, s = 1, \ldots, k$, where, $N_l/N \rightarrow \tilde{s}_l$ and for $r, l = 1, \dots, k$,
\begin{equation*}
\label{eq:Ys}
Y_i^{(r,l)} \equiv p_r(\tilde{X}_i^{(l)}, \bzeta^0) - E_{q_l} \bigl( p_r(X, \bzeta^0) \bigr), \qquad i = 1, \ldots, N_l.
\end{equation*}
As in \cite{roy:tan:fleg:2018}, this will be proved in couple of
  steps. First, we consider a single chain $\tilde{\Phi}_l$ used to calculate
  $k$ quantities. We use the results
  in \citet{vats:fleg:jone:2015} who obtain conditions for the
  multivariate SV estimator to be strongly consistent. Second, we combine results from the $k$
  independent chains.  Finally, we show that $\widehat{\Omega}$ is a
  strongly consistent estimator of $\Omega$.

  Denote
  $\bar{Y}^{(l)} = \left( \bar{Y}^{(1,l)}, \bar{Y}^{(2,l)}, \dots,
    \bar{Y}^{(k,l)} \right)^{\top}$ where
  $\bar{Y}^{(r,l)} = \sum_{i=1}^{N_l} Y_i^{(r,l)}/N_l$. From
  \cite{roy:tan:fleg:2018} we have
  $\sqrt{N_l}\bar{Y}^{(l)} \cd {\cal N} (0, \Sigma^{(l)})$ as
  $N_l \rightarrow \infty$, where $\Sigma^{(l)}$ is a $k \times k$
  covariance matrix with
\begin{equation}\label{eq:Sigmars}
 \Sigma^{(l)}_{rs} = E_{q_l}\{Y_1^{(r,l)} Y_1^{(s,l)}\} +  \sum_{i=1}^{\infty} E_{q_l}\{Y_1^{(r,l)} Y_{1+i}^{(s,l)}\}+  \sum_{i=1}^{\infty} E_{q_l}\{Y_{1+i}^{(r,l)} Y_1^{(s,l)}\} .
\end{equation}

The SV estimator of $\Sigma^{(l)}$ is given in (3.11). 
We now prove the strong consistency of
$\widehat{\Sigma}^{(l)}$. Note that $\widehat{\Sigma}^{(l)}$ is
defined using the terms $\bar{Z}^{(l)}_{i}$'s which involve the
random quantity $\hatbzeta$. We define $\widehat{\Sigma}^{(l)}
(\bzeta^0)$ to be $\widehat{\Sigma}^{(l)}$ with $\bzeta^0$ substituted
for $\hatbzeta$, that is,
\[
\widehat{\Sigma}^{(l)} (\bzeta^0)
 =   \frac{1}{N_l} \sum_{j = -(b_{N_l} -1)}^{b_{N_l} -1} w_{N_l}(j) \sum_{i \in S_{j,N}} \left[ Y^{(l)}_{i} - \bar{Y}^{(l)} \right] \left[ Y^{(l)}_{i+j} - \bar{Y}^{(l)} \right]^{\top}  \mbox{ for} \;\; l =1,\ldots, k,
\]
where $Y^{(l)}_{i}= \left( Y^{(1,l)}_{i}, \ldots,
  Y^{(k,l)}_{i} \right) ^{\top}$. We prove $\widehat{\Sigma}^{(l)} \cas
\Sigma^{(l)}$ in two steps: (1) $\widehat{\Sigma}^{(l)} (\bzeta^0) \cas\Sigma^{(l)}$ and (2) $\widehat{\Sigma}^{(l)} - \widehat{\Sigma}^{(l)} (\bzeta^0) \cas 0$. Under the conditions of Theorem 1,
it follows from \citet{vats:fleg:jone:2015} that
$\widehat{\Sigma}^{(l)} (\bzeta^0) \cas \Sigma^{(l)}$ as
$N_l \rightarrow \infty$. We show
$\widehat{\Sigma}_{rs}^{(l)} - \widehat{\Sigma}_{rs}^{(l)} (\bzeta^0)
\cas 0$ where $\widehat{\Sigma}_{rs}^{(l)}$ and
$\widehat{\Sigma}_{rs}^{(l)} (\bzeta^0)$ are the $(r,s)$th elements of
the $k \times k$ matrices $\widehat{\Sigma}_{rs}^{(l)}$ and
$\widehat{\Sigma}_{rs}^{(l)} (\bzeta^0)$ respectively. By the mean
value theorem (in multiple variables), there exists
$\bzeta^*=t\hatbzeta +(1-t) \bzeta^0$ for some $t\in(0,1)$, such that
\begin{equation}
  \label{eq:Sigmv}
\widehat{\Sigma}_{rs}^{(l)} - \widehat{\Sigma}_{rs}^{(l)} (\bzeta^0) =
\nabla \widehat{\Sigma}_{rs}^{(l)} (\bzeta^*) \cdot (\hatbzeta -
\bzeta^0),
\end{equation}
 where
$\cdot$ represents the dot product. Note that
 \[
 \widehat{\Sigma}_{rs}^{(l)} (\bzeta) = \frac{1}{N_l} \sum_{j = -(b_{N_l} -1)}^{b_{N_l} -1} w_{N_l}(j) \sum_{i} \left[ Z^{(r,l)}_{i}(\bzeta) - \bar{Z}^{(r,l)} (\bzeta)\right] \left[ Z^{(s,l)}_{i+j} (\bzeta)- \bar{Z}^{(s,l)}(\bzeta) \right],
 \]
 where $Z^{(r,l)}_{i}(\bzeta) := p_r(\tilde{X}_i^{(l)}, \bzeta)$ and $\bar{Z}^{(r,l)} (\bzeta) := \sum_{j=1}^{N_l}
 p_r(\tilde{X}_j^{(l)}, \bzeta)/N_l$. Some calculations show that for $t \neq r$
 \[
 \frac{\partial Z^{(r,l)}_{j} (\bzeta)}{\partial \bzeta_t} = - p_r(\tilde{X}_j^{(l)}, \bzeta) p_t(\tilde{X}_j^{(l)}, \bzeta)
 \]
 and
 \[
 \frac{\partial Z^{(r,l)}_{j} (\bzeta)}{\partial \bzeta_r} = p_r(\tilde{X}_j^{(l)}, \bzeta) (1 - p_r(\tilde{X}_j^{(l)}, \bzeta)).
 \]
 Simplifying the notations, we denote
 $U^{(r,t)}_j := \partial Z^{(r,l)}_{j} (\bzeta)/\partial \bzeta_t$,
 $\bar{U}^r := \partial \bar{Z}^{(r,l)} (\bzeta)/\partial \bzeta_t$
 and simply write $Z^{(r,l)}_{j}$ and $\bar{Z}^{(r,l)}$ for
 $Z^{(r,l)}_{j} (\bzeta)$ and $\bar{Z}^{(r,l)} (\bzeta)$ respectively.  Thus we
 have
\begin{align}
  & \frac{\partial  \widehat{\Sigma}_{rs}^{(l)} (\bzeta) }{\partial \bzeta_t} =
    \frac{1}{N_l} \sum_{j = -(b_{N_l} -1)}^{b_{N_l} -1} w_{N_l}(j) \sum_{i} \left[
    (Z^{(r,l)}_{i} - \bar{Z}^{(r,l)}) (U^{(s,t)}_{i+j} - \bar{U}^{(s,t)}) +  (U^{(r,t)}_i - \bar{U}^{(r,t)}) (Z^{(s,l)}_{i+j} - \bar{Z}^{(s,l)})\right]\nonumber\\
  =& \frac{1}{N_l} \sum_{j = -(b_{N_l} -1)}^{b_{N_l} -1} w_{N_l}(j) \sum_{i} \left[
    (Z^{(r,l)}_{i} - \bar{Z}^{(r,l)}) (U^{(s,t)}_{i+j} - \bar{U}^{(s,t)}) \right] \label{eq:delSig}\\ & \hspace{1in} + \frac{1}{N_l} \sum_{j = -(b_{N_l} -1)}^{b_{N_l} -1} w_{N_l}(j) \sum_{i} \left[(U^{(r,t)}_i - \bar{U}^{(r,t)}) (Z^{(s,l)}_{i+j} - \bar{Z}^{(s,l)})\right]\label{eq:delSig1},
\end{align}
Let $V_i^{(l)} := (Z^{(r,l)}_{i}, U^{(s,t)}_i)^T$ and
\[
\widehat{\Sigma}^{(l)}_V (\bzeta)
 =   \frac{1}{N_l} \sum_{j = -(b_{N_l} -1)}^{b_{N_l} -1} w_{N_l}(j) \sum_{i} \left[ V^{(l)}_{i} - \bar{V}^{(l)} \right] \left[ V^{(l)}_{i+j} - \bar{V}^{(l)} \right]^{\top} .
\]
Since $p_r(X, \bzeta)$ is uniformly bounded by 1 and $\tilde{\Phi}_l$ is
polynomially ergodic of order $m > 1$, from
\citet{vats:fleg:jone:2015} we know that
$\widehat{\Sigma}^{(l)}_V (\bzeta) \cas \Sigma^{(l)}_V (\bzeta)$ where
$\Sigma^{(l)}_V (\bzeta)$ is the covariance matrix of the asymptotic
distribution of $\sqrt{N_l} (\bar{V}^{(l)} - E_{q_l} V)$. Since the
expression in \eqref{eq:delSig} is the off-diagonal elements of
$\widehat{\Sigma}^{(l)}_V (\bzeta)$, it is bounded with probability
one. We can similarly see that the expression in \eqref{eq:delSig1} is
bounded with probability one. Note that, the proof to show that
$\partial \widehat{\Sigma}_{rs}^{(l)} (\bzeta)/\partial \bzeta_t$ is
bounded with probability one is quite different from the proof in
\cite{roy:tan:fleg:2018}.

Note that the terms $Z^{(r,l)}_{i}, U_{i}^{(r,t)}$, etc, above
actually depends on $\bzeta$, and we are indeed concerned with the
case where $\bzeta$ takes on the value $\bzeta^*$, lying between
$\hatbzeta$ and $\bzeta^0$. Since, $\hatbzeta \cas \bzeta^0$, we have
$\bzeta^* \cas \bzeta^0$ as $N_l \rightarrow \infty$. Let $\|u \|_{L_1}$
denotes the $L_1$ norm of a vector $u \in \mathbb{R}^k$. So from
\eqref{eq:Sigmv}, and the fact that
$\partial \widehat{\Sigma}_{rs}^{(l)} (\bzeta)/\partial \bzeta_t$ is
bounded with probability one, we have
  \[
    |\widehat{\Sigma}_{rs}^{(l)} - \widehat{\Sigma}_{rs}^{(l)} (\bzeta^0)| \leq \underset{1 \le t \le k}{\max}\left\{\left|\frac{\partial  \widehat{\Sigma}_{rs}^{(l)} (\bzeta^*) }{\partial \bzeta_t}\right|\right\} \|\hatbzeta - \bzeta^0\|_{L_1} \cas 0
    \;\;\text{ as}\; n\rightarrow \infty.
 \]
Let
\begin{equation*}
  \label{eq:Sigmahat}
\widehat{\Sigma} = \begin{pmatrix}
\widehat{\Sigma}^{(1)}\Ddot{11}.(6pt,-2pt,6pt,-1.5pt)&\quad&\quad&\quad 0&\\
&&&&\\
&0&&&\widehat{\Sigma}^{(k)}\\
\end{pmatrix} .
\end{equation*}

 Since $\widehat{\Sigma}^{(l)} \cas \Sigma^{(l)},$ for
 $l = 1,\dots,k$, it follows that $\widehat{\Sigma} \cas \Sigma$
where
$\Sigma$ is the corresponding $k^2 \times k^2$ covariance matrix, that
is, $\Sigma$ is a block diagonal matrix as $\widehat{\Sigma}$ with
$\Sigma^{(l)}$ substituted for $\widehat{\Sigma}^{(l)}, l=1,\dots,k$.
Define the following $k \times k^2$ matrix
\begin{equation*}
  \label{eq:defA}
A_N = \left( - \sqrt{\frac{N}{N_1}} a_1 I_k \quad - \sqrt{\frac{N}{N_2}} a_2 I_k \quad \dots \quad - \sqrt{\frac{N}{N_k}} a_k I_k \right) \;,
\end{equation*}
where $I_k$ denotes the $k \times k$ identity matrix.
Then we have
$\widehat{\Omega} \equiv A_N \widehat{\Sigma} A_N^{T} \cas \Omega$ as
$N \rightarrow \infty$.

\section{Proof of Theorem 2 (a)}
\label{sec:appthm2}
From
\cite{roy:tan:fleg:2018} we know that
$ \sigma^2_u = \varpi c(\pi; \bd)^{\top} V c(\pi; \bd) + \tau^2 (\pi ;
\bd)$, where $\tau^2 (\pi ;  \bd) = \sum_{l =1}^k (a_l^2/s_l) \tau^2_l(\pi ;
 \bd)$, and
 \begin{equation}
   \label{eq:taul}
  \tau^2_l(\pi ; \bd) = \mbox{Var}_{q_l} (u^{\pi}(X_1^{(l)}; \bd)) + 2 \sum_{g=1}^{\infty}
\mbox{Cov}_{q_l} (u^{\pi}(X_1^{(l)};  \bd), u^{\pi}(X_{1+g}^{(l)};  \bd)).
\end{equation}
To prove Theorem 2~(a), note that, we already have a
consistent SV estimator $\widehat{V}$ of $V$. From
\cite{roy:tan:fleg:2018} it follows that
$\hat{c}(\pi;  \hatbd)^{\top} \widehat{V} \hat{c}(\pi;
\hatbd) \cas c(\pi;  \bd)^{\top} V c(\pi;  \bd)$.

We now show $\hat{\tau}_l^2 (\pi ;  \hatbd)$ is a
consistent estimator of $\tau_l^2 (\pi ;  \bd)$ where
 $\hat{\tau}^2_l$ is defined in~(3.19). 
Since the Markov chains
$\{X_i^{(l)}\}_{i=1}^{n_l}, l=1,\dots,k$ are independent, it
then follows that $\tau^2 (\pi ;  \bd)$ is consistently estimated
by $\hat{\tau}^2 (\pi ;  \hatbd)$ completing the proof of
Theorem 2~(a).

If $\bd$ is known from the assumptions of Theorem 2 
and the results in \citet{vats:fleg:jone:2015}, 
we know that $\tau_l^2 (\pi ;  \bd)$ is
consistently estimated by its SV estimator $\hat{\tau}_l^2
(\pi ;  \bd)$. Note that, $\hat{\tau}_l^2 (\pi ;  \bd)$ is
defined in terms of the quantities $u^{\pi}(X_i^{(l)};  \bd)$'s. We now
show that $\hat{\tau}_l^2 (\pi ;  \hatbd) - \hat{\tau}_l^2 (\pi ;
 \bd) \cas 0.$
Let
\[
{\partial U} _i^m( \bz) := \frac{\partial u_{i}( \bz)}{\partial z_m}
  =\frac{a_m}{ z_m^2} \frac{\nu(X_i^{(l)}) \varphi_{m}(X_i^{(l)})}{\left(\sum_s a_s \varphi_{s}(X_i^{(l)})/z_s \right)^2},
\]
and $\bar{\partial U}^m(\bz)$ be the averages of $\{{\partial U} _i^m( \bz), i =1, \dots, n_l\}$.
Denoting $\hat{\tau}_l^2 (\pi ;  \bz)$ by $G(\bz)$, by the mean value theorem (in multiple variables), there exists $\bd^*=t\hatbd+(1-t)\bd$ for some $t\in(0,1)$, such that $G(\hatbd)-G(\bd) = \nabla G(\bd^*) \cdot (\hatbd-\bd)$.  For any $m \in \{2,\cdots, k\}$, and $\bz \in {R^+}^{k-1}$,
\begin{align*}
& \frac{\partial G(\bz)}{\partial z_m}
= \sum_{j = -(b_{n_l} -1)}^{b_{n_l} -1} w_{n_l}(j) \sum_{i} \left[u_{i}( \bz) - \bar{u}(
    \bz)\right] \left[ {\partial U}^m_{i+j}( \bz) - \bar{\partial U}^m(
    \bz)\right]\\
& \hspace{.6in}+\sum_{j = -(b_{n_l} -1)}^{b_{n_l} -1} w_{n_l}(j) \sum_{i} \left[{\partial U}^m_{i}( \bz) - \bar{\partial U}^m(
    \bz)\right] \left[ u_{i+j}( \bz) - \bar{u}(
    \bz)\right]
\end{align*}
Then using similar arguments as in the proof of Theorem 1, it can be shown that
$\partial G(\bz)/\partial z_m$ is bounded with probability one. Then it follows that
 \begin{equation*}
 |G(\hatbd)-G(\bd)| \leq  \underset{1 \le m \le k-1}{\max} \left\{\left| \frac{\partial G(\bd^*)}{\partial z_m}\right|\right\} \|\hatbd-\bd\|_{L_1} \cas 0.
 \end{equation*}

\section{Proof of Theorem 2 (b)}
\label{sec:appthm3}
From
\cite{roy:tan:fleg:2018} we know that
$\sigma^2_{\eta} = \varpi
e(\pi;  \bd)^{\top} V e(\pi;  \bd) + \rho (\pi ;  \bd)$, where
\begin{equation*}
  \label{eq:defrho}
  \rho(\pi;  \bd) = \nabla h(E_\pi f u(\pi, \pi_1), u(\pi, \pi_1))^{\top} \Gamma(\pi;  \bd) \nabla h( E_\pi f u(\pi, \pi_1), u(\pi, \pi_1)),
\end{equation*}
\begin{equation*}
\Gamma (\pi;  \bd) = \sum_{l =1}^k \frac{a_l^2}{s_l} \Gamma_l(\pi;  \bd);     \Gamma_l(\pi;  \bd) = \left(
  \begin{array}{lr}
    \gamma^{11} & \gamma^{12}\\
    \gamma^{21} & \gamma^{22}\\
  \end{array}
\right),
\end{equation*}
with
\begin{align*}
  \gamma^{11} \equiv \gamma^{11}_l(\pi; \bd)
  &= \mbox{Var}_{q_l} (v^{[f], \pi}(X_1^{(l)};  \bd)) + 2 \sum_{g=1}^{\infty}
    \mbox{Cov}_{q_l} (v^{[f], \pi}(X_1^{(l)};  \bd), v^{[f], \pi}(X_{1+g}^{(l)};
     \bd)), \\
  \gamma^{12} \equiv \gamma^{12}_l(\pi; \bd) &= \gamma^{21}
  \equiv \gamma^{21}_l(\pi; \bd) \\
  &= \mbox{Cov}_{q_l} (v^{[f], \pi}(X_1^{(l)};  \bd), u^{\pi}(X_1^{(l)};
    \bd)) \\
  &\quad{}+ \sum_{g=1}^{\infty} [\mbox{Cov}_{q_l} (v^{[f], \pi}(X_1^{(l)};
    \bd), u^{\pi}(X_{1+g}^{(l)};  \bd)) +\mbox{Cov}_{q_l}
    (v^{[f], \pi}(X_{1+g}^{(l)};  \bd), u^{\pi}(X_1^{(l)};  \bd))] \\
  \gamma^{22}_l \equiv\gamma^{22}_l(\pi;  \bd)
  &= \mbox{Var}_{q_l} (u^{\pi}(X_1^{(l)};  \bd)) + 2 \sum_{g=1}^{\infty}
    \mbox{Cov}_{q_l} (u^{\pi}(X_1^{(l)};  \bd), u^{\pi}(X_{1+g}^{(l)};  \bd)) .
\end{align*}
From \cite{roy:tan:fleg:2018} we know that $\hat{e}(\pi;  \hatbd)^{\top} \widehat{V} \hat{e}(\pi;
\hatbd) \cas e(\pi;  \bd)^{\top} V e(\pi;  \bd)$. Thus,
to prove Theorem 2~(b),  we only
need to show that
$\widehat{\Gamma}_l(\pi; \hatbd) \cas \Gamma_l(\pi; \bd)$. Note that,
 \begin{align*}
  \widehat{\Gamma}_{l} (\pi ;  \bd)
&= \frac{1}{n_l} \sum_{j = -(b_{n_l} -1)}^{b_{n_l} -1} w_{n_l}(j) \sum_{i \in S_{j,n}}
 \left(
  \begin{array}{cc}
   \left[v^{[f]}_{i} - \bar{v}^{[f]}\right] \left[v^{[f]}_{i+j} - \bar{v}^{[f]}\right]&  \left[v^{[f]}_{i} - \bar{v}^{[f]}\right]\left[u_{i+j} - \bar{u}\right] \\
 \left[ v^{[f]}_{i+j} - \bar{v}^{[f]}\right]\left[ u_{i} - \bar{u}\right]  & \left[u_{i} - \bar{u}\right] \left[u_{i+j} - \bar{u}\right]\\
 \end{array}
 \right).
  \\
&=
\left(\begin{array}{cc}
    \hat{\gamma}^{11}(\pi ;  \bd) & \hat{\gamma}^{12}(\pi ;  \bd)\\
    \hat{\gamma}^{21}(\pi ;  \bd) & \hat{\gamma}^{22}(\pi ;  \bd)\\
\end{array}
\right).
      \end{align*}

If $\bd$ is
known, from the assumptions of Theorem~2~(b) and the results
in \citet{vats:fleg:jone:2015}, we know that $\Gamma_l (\pi ; \bd)$ is
consistently estimated by its SV estimator
$\widehat{\Gamma}_l (\pi ; \bd)$.  We now show that
$\widehat{\Gamma}_l (\pi ; \hatbd) - \widehat{\Gamma}_l (\pi ; \bd)
\cas 0.$

From Theorem~2~(a), we know that
$\hat{\gamma}_l^{22} (\pi ; \hatbd) - \hat{\gamma}_l^{22} (\pi ; \bd)
\cas 0$ as $\gamma^{22}_l$ is the same as $\tau^2_l(\pi; \bd)$
defined in \eqref{eq:taul}. We now show
$\hat{\gamma}_l^{11} (\pi ; \hatbd) - \hat{\gamma}_l^{11} (\pi ;
\bd)\cas 0$.

Let
\[
{\partial V} _i^{[f],m}( \bz) := \frac{\partial v^{[f]}_{i}( \bz)}{\partial z_m}
  =\frac{a_m}{ z_m^2} \frac{f(X_i^{(l)}) \nu(X_i^{(l)}) \varphi_{m}(X_i^{(l)})}{\left(\sum_s a_s \varphi_{s}(X_i^{(l)})/z_s \right)^2},
\]
and $\bar{\partial V}^{[f],m}(\bz)$ be the averages of $\{{\partial V}_i^{[f],m}( \bz), i =1, \dots, n_l\}$.

Letting $\hat{\gamma}_l^{11} (\pi ;  \bz)$ by $H(\bz)$,
 by the mean value theorem, there exists
$\bd^*=t\hatbd+(1-t)\bd$ for some $t\in(0,1)$, such that $H(\hatbd)-H(\bd) = \nabla H(\bd^*) \cdot (\hatbd-\bd)$.  For any $m \in \{2,\cdots, k\}$, and $\bz \in {R^+}^{k-1}$,
\begin{align*}
& \frac{\partial H(\bz)}{\partial z_m}
= \sum_{j = -(b_{n_l} -1)}^{b_{n_l} -1} w_{n_l}(j) \sum_{i} \left[v^{[f]}_{i}( \bz) - \bar{v}^{[f]}(
    \bz)\right] \left[ {\partial V}^{[f],m}_{i+j}( \bz) - \bar{\partial V}^{[f], m}(
    \bz)\right]\\
& \hspace{.6in} +\sum_{j = -(b_{n_l} -1)}^{b_{n_l} -1} w_{n_l}(j) \sum_{i} \left[{\partial V}^{[f], m}_{i}( \bz) - \bar{\partial V}^{[f],m}(
    \bz)\right] \left[v^{[f]}_{i+j}( \bz) - \bar{v}^{[f]}(
    \bz)\right]
\end{align*}
The rest of the proof is analogous to Theorem~2~(a) and thus
we have
$\hat{\gamma}_l^{11} (\pi ;  \hatbd) - \hat{\gamma}_l^{11} (\pi ;
 \bd)\cas 0$. Finally, using similar arguments as before we can show
$\hat{\gamma}_l^{12} (\pi ;  \hatbd) - \hat{\gamma}_l^{12} (\pi ; \bd)\cas 0$.

\section{Proof of Theorem 3}
\label{sec:appthm4}
Since the Markov chains used in stage 1 are polynomially ergodic of
order $m > 1$, from \citet[][Theorem 1]{roy:tan:fleg:2018}, we have
$N^{1/2} (\hatbd_{\bq} - \bd_{\bq}) \cd {\cal N} (0,
V_{\bq})$. Since $n/N \rightarrow r$, it follows that
$\sqrt{n} (\hatbd_{\bq} - \bd_{\bq}) \cd {\cal N} (0, \varpi
V_{\bq})$. Following \citet[][Proof of Theorem 2]{roy:tan:fleg:2018}
we write
\begin{equation}
  \label{eq:huminu}
  \begin{split}
      \sqrt{n}(\hat{\bf u}(\bpi; \hatbd_{\bq}) - {\bf u}(\bpi, q_1)) = \sqrt{n}(\hat{\bf u}(\bpi; \hatbd_{\bq}) - \hat{\bf u}(\bpi; \bd_{\bq})) + \sqrt{n}(\hat{\bf u}(\bpi; \bd_{\bq}) - {\bf u}(\bpi, q_1)).
  \end{split}
\end{equation}
Note that the 2nd term involves randomness only from the
2nd stage Markov chains. Since $\sum_{l=1}^k a_l E_{q_l}{\bf u}^{\bpi}(X; \bd_{\bq}) = {\bf u}(\bpi, q_1)$,
we have
 \[
   \sqrt{n}(\hat{\bf u}(\bpi;\bd_{\bq}) - {\bf u}(\bpi,
   q_1)) = \sum_{l=1}^k a_l \sqrt{\frac{n}{n_l}}
   \frac{\sum_{i=1}^{n_l} ({\bf u}^{\bpi}(X_i^{(l)}; \bd_{\bq}) - E_{q_l} {\bf u}^{\bpi}(X;
     \bd_{\bq}))}{\sqrt{n_l}}.
\]
Since $\Phi_l$ is polynomially ergodic of order $m$ and
$E_{q_l} |u^{\pi}(X; \bd_{\bq})|^{2+\delta}$ is finite for each
$\pi \in \Pi$ where $m > 1 + 2/\delta$, it
follows that
$\sum_{i=1}^{n_l} ({\bf u}^{\bpi}(X_i^{(l)}; \bd_{\bq}) - E_{q_l} {\bf u}^{\bpi}(X;
\bd_{\bq}))/\sqrt{n_l} \cd N(0, T_l(\bd_{\bq}))$ where $T_l(\bd_{\bq})$
is the matrix with elements defined in~(3.20). 
As $n_l/n \rightarrow s_l$ and the Markov chains $\Phi_l$'s are
independent, it follows that
$ \sqrt{n}(\hat{\bf u}(\bpi;
\bd_{\bq}) - {\bf u}(\bpi, q_1))
\cd N(0, \sum_{l=1}^k (a_l^2/s_l) T_l(\bd_{\bq}))$.

Next by Taylor series
expansion of $F(\bd) \equiv \hat{u}(\pi; \bd)$ about $\bd_{\bq}$, we have
\begin{equation*}
  \label{eq:Ftayl}
  \sqrt{n}(F(\hatbd_{\bq}) - F(\bd_{\bq})) = \sqrt{n} \nabla F(\bd_{\bq})^{\top} (\hatbd_{\bq} - \bd_{\bq}) + \frac{\sqrt{n}}{2} (\hatbd_{\bq} - \bd_{\bq})^{\top} \nabla^2 F(\bd^*) (\hatbd_{\bq} - \bd_{\bq}),
\end{equation*}
where $\bd^*$ is between $\bd_{\bq}$ and $\hatbd_{\bq}$. As in
\cite{roy:tan:fleg:2018}, we can then show that
\[
\sqrt{n}(\hat{u}(\pi; \hatbd_{\bq}) - \hat{u}(\pi; \bd_{\bq})) = \sqrt{q} c(\pi; \bd_{\bq}) \sqrt{N} (\hatbd_{\bq} - \bd_{\bq}) + o_p(1) .
\]
Accumulating the terms for all $\pi \in \Pi$, we have
\[
\sqrt{n}(\hat{\bf u}(\bpi; \hatbd_{\bq}) - \hat{\bf u}(\bpi; \bd_{\bq})) = \sqrt{q} C(\bpi; \bd_{\bq}) \sqrt{N} (\hatbd_{\bq} - \bd_{\bq}) + {\bf o}_p(1) .
\]
Thus for constant vectors $t_1$ and $t_2$ of dimensions $k-1$ and
$|\Pi|$ respectively, we have
\begin{align}
\label{eq:crwo}
 & t_1^{\top} \sqrt{n} (\hatbd_{\bq} - \bd_{\bq}) + t_2^{\top} \sqrt{n} (\hat{\bf u}(\bpi; \hatbd_{\bq}) - {\bf u}(\bpi, q_1)) \nonumber\\ &= \sqrt{\varpi} (t_1^{\top} + t_2^{\top}C(\bpi; \bd_{\bq})) \sqrt{N} (\hatbd_{\bq} - \bd_{\bq}) + \sum_{l=1}^k a_l \sqrt{\frac{n}{n_l}}
   \frac{\sum_{i=1}^{n_l} t_2^{\top} ({\bf u}^{\bpi}(X_i^{(l)}; \bd_{\bq}) - E_{q_l} {\bf u}^{\bpi}(X;
     \bd_{\bq}))}{\sqrt{n_l}} + o_p(1) \nonumber\\
&\cd N(0, \varpi(t_1^{\top} + t_2^{\top}C(\bpi; \bd_{\bq}))V_{\bq} (t_1 +C(\bpi; \bd_{\bq})^{\top}t_2) +
\sum_{l=1}^k (a_l^2/s_l) t_2^{\top} T_l(\bd_{\bq})t_2),
\end{align}
where the last step follows from the independence of the Markov chains
involved in the two stages. Note that the variance in \eqref{eq:crwo}
is the same as
\[ (t_1^{\top}, t_2^{\top})\left(\begin{array}{cc}
                                                 \varpi V_{\bq} & \Sigma_{12}\\
                                                 \Sigma_{21} & \Sigma_{22}\\
\end{array} \right) (t_1^{\top}, t_2^{\top})^{\top}.
\]
Hence the Cram\'{e}r-Wold device implies the joint central limit theorem (CLT) in (3.22). 
Thus Theorem~3~(a) is proved.

From the proofs of Theorems~1 and~2~(a), we know that
$\varpi \widehat{C}(\bpi; \hatbd_{\bq})
\widehat{V}_{\bq} \widehat{C}(\bpi;
\hatbd_{\bq})^{\top}$ is a consistent estimator of
$\varpi C(\bpi; \bd_{\bq})
V_{\bq} C(\bpi;
\bd_{\bq})^{\top}$. If $\bd_{\bq}$ is known from the assumptions of
Theorem~3~(b) and the results in \citet{vats:fleg:jone:2015}, we know that
$T_l (\bd_{\bq})$ is consistently estimated by its SV estimator
$\widehat{T}_l (\bd_{\bq})$ defined in~(3.21). 
 Then using
similar arguments as in the proof of Theorem~2~(a), we can
show that every element of
$\widehat{T}_l (\bd_{\bq})- \widehat{T}_l (\hatbd_{\bq})$ converges to
zero (a.e.). Hence Theorem~3~(b) is proved.

\section{Entropy decomposition for multiple IS estimators of means}
\label{sec:appthm5}
In this section, we prove a result similar to Theorem~3 for
$\hat{\eta}^{[f]} (\pi; \hatbd)$. Let
$\hat{\bbeta}^{[f]}(\bpi; \hatbd_{\bq})$
be the vector of length $|\Pi|$ consisting of
$\hat{\eta}^{[f]}(\pi; \hatbd_{\bq})$'s,
$\pi \in \Pi$ in a fixed order. Similarly
define $\hat{\bf v}^{[f]}(\bpi; \hatbd_{\bq})$ and the vector of true means
${\bf E}_{\bpi} f$. Let
$p^* \equiv |\Pi|$. Let
$E(\bpi; \bd_{\bq})$ be the
$p^* \times (k-1)$ matrix with rows $e(\pi; \bd_{\bq})$ (defined in
Section~3.2 of the paper), $\pi \in \Pi$. Similarly, define
$\widehat{E}(\bpi; \bd_{\bq})$ with rows
$\hat{e}(\pi; \bd_{\bq})$, $\pi \in \Pi$. Let ${\bf v}^{[f], \bpi} (x ;\bd_{\bq})$ be
the $p$ dimensional vector consisting of $v^{[f], \pi} (x; \bd_{\bq})$'s defined in (3.14) of the paper,
$\pi \in \Pi$. Define the $2p^* \times 2p^*$ matrix
\begin{equation}
  \label{eq:lamd}
  \Lambda_l(\bd_{\bq}) = \left(\begin{array}{cc}
   \Lambda_l^{11}(\bd_{\bq})  & \Lambda_l^{12}(\bd_{\bq})\\
    \Lambda_l^{21}(\bd_{\bq}) & T_l (\bd_{\bq})\\
\end{array} \right),
\end{equation}
where the
elements of $\Lambda_l^{11}(\bd_{\bq})$ are given by
\begin{align*}
  \lambda^{11}_l(\pi, \pi' ; \bd_{\bq}) = \mbox{Cov}_{q_l} (v^{[f],\pi}(X_1^{(l)}; \bd_{\bq}), v^{[f],\pi'}(X_1^{(l)}; \bd_{\bq})) & + \sum_{g=1}^{\infty}
\mbox{Cov}_{q_l} (v^{[f],\pi}(X_1^{(l)};  \bd_{\bq}), v^{[f],\pi'}(X_{1+g}^{(l)};  \bd_{\bq}))\\  & + \sum_{g=1}^{\infty}
\mbox{Cov}_{q_l} (v^{[f],\pi}(X_{1+g}^{(l)};  \bd_{\bq}), v^{[f],\pi'}(X_{1}^{(l)};  \bd_{\bq})),
\end{align*}
and the
elements of $\Lambda_l^{12}(\bd_{\bq})$ are given by
\begin{align*}
  \label{eq:gamxiluv}
  \lambda^{12}_l(\pi, \pi' ; \bd_{\bq}) = \mbox{Cov}_{q_l} (v^{[f],\pi}(X_1^{(l)}; \bd_{\bq}), u^{\pi'}(X_1^{(l)}; \bd_{\bq})) & + \sum_{g=1}^{\infty}
\mbox{Cov}_{q_l} (v^{[f],\pi}(X_1^{(l)};  \bd_{\bq}), u^{\pi'}(X_{1+g}^{(l)};  \bd_{\bq}))\\  & + \sum_{g=1}^{\infty}
\mbox{Cov}_{q_l} (v^{[f],\pi}(X_{1+g}^{(l)};  \bd_{\bq}), u^{\pi'}(X_{1}^{(l)};  \bd_{\bq})).
\end{align*}
 Also $\Lambda_l^{21}(\bd_{\bq}) = \Lambda_l^{12}(\bd_{\bq})^{\top}$ and let $ \Lambda (\bd_{\bq}) = \sum_{l=1}^k (a_l^2/s_l)\Lambda_l(\bd_{\bq})$.
Define a function ${\bf h} : \mathbb{R}^{2p^*} \rightarrow \mathbb{R}^{p^*}$ where
\[
{\bf h} (x_1,\dots, x_{2p^*}) = \Big(\frac{x_1}{x_{p^*+1}}, \frac{x_2}{x_{p^*+2}}, \dots, \frac{x_{p^*}}{x_{2p^*}}\Big)
\]
with its gradient given by
\[
  \nabla {\bf h} (x) =
  \begin{pmatrix}
    1/x_{p^*+1}    & 0    & \ldots &0 & -x_1/x^2_{p^*+1} & 0 &\ldots &0   \\
    0   & 1/x_{p^*+2}      & \ldots &0 & 0  & -x_2/x^2_{p^*+2} &\ldots&0    \\
      \vdots & \vdots &\vdots &\vdots &\vdots &\vdots & \ddots & \vdots \\
    0  &0      & \ldots & 1/x_{2p^*}&0&0&\ldots&-x_{p^*}/x^2_{2p^*}
  \end{pmatrix}.
\]
Define the $p^* \times p^*$ matrix
\[
\boldsymbol{\rho} (\bd_{\bq}) = \nabla {\bf h}({\bf E}_{\bpi} f \odot {\bf u}(\bpi, q_1), {\bf u}(\bpi, q_1))^{\top} \Lambda(\bd_{\bq})\nabla {\bf h}({\bf E}_{\bpi} f \odot {\bf u}(\bpi, q_1), {\bf u}(\bpi, q_1)^{\top},
\]
where $\odot$ denotes element-wise multiplication.
Let
\begin{equation}
  \label{eq:gamxildef}
  \widehat{\Lambda}_l(\bd_{\bq})= \frac{1}{n_l} \sum_{j = -(b_{n_l} -1)}^{b_{n_l} -1} w_{n_l}(j) \sum_{i \in S_{j,n}} \left(\begin{array}{c}
   {\bf v}^{[f],\bpi}(X_{i}^{(l)}; \bd_{\bq}) - \bar{{\bf v}^{[f]}}(\bd_{\bq})\\
    {\bf u}^{\bpi}(X_{i}^{(l)}; \bd_{\bq}) - \bar{{\bf u}}(\bd_{\bq})\\
\end{array} \right) \left(\begin{array}{c}
   {\bf v}^{[f],\bpi}(X_{i+j}^{(l)}; \bd_{\bq}) - \bar{{\bf v}^{[f]}}(
    \bd_{\bq})\\
    {\bf u}^{\bpi}(X_{i+j}^{(l)}; \bd_{\bq}) - \bar{{\bf u}}(
    \bd_{\bq})\\
\end{array} \right)^{\top},
\end{equation}
where $b_{n_l}$'s are the truncation points, $w_{n_l}(j)$'s are lag
window, and
$\bar{\bf v}^{[f]}( \bd_{\bq})= \sum_{i=1}^{n_l}{\bf v}^{[f], \bpi}(X_{i}^{(l)};  \bd_{\bq})/n_l$. Let $\widehat{\Lambda}(\bd_{\bq}) \equiv \sum_{l=1}^k (a_l^2n/n_l) \widehat{\Lambda}_l(\bd_{\bq})$. Finally, let
\[
\boldsymbol{\hat{\rho}} (\hatbd_{\bq}) = \nabla {\bf h}(\hat{\bf v}^{[f]}(\bpi; \hatbd_{\bq}), \hat{\bf u}(\bpi;
\hatbd_{\bq})) \widehat{\Lambda}(\hatbd_{\bq})\nabla {\bf h}(\hat{\bf v}^{[f]}(\bpi; \hatbd_{\bq}),\hat{\bf u}(\bpi; \hatbd_{\bq}))^{\top},
\]
\noindent {\bf Theorem 4}
{\it Suppose that $N_l, n_l \rightarrow \infty$ for all
  $l= 1,\ldots,k$, and there exists $\varpi \in [0, \infty)$ such that
  $n/N \rightarrow \varpi$. Here, $N\equiv \sum_{l=1}^k$ and
  $n=\sum_{l=1}^k n_l$ are the total sample sizes for stages 1 and 2,
  respectively. In addition, let $n_l/n \rightarrow s_l$ for
  $l = 1,\cdots,k$.
\begin{enumerate}
\item[(a)] Assume that the stage 1 Markov chains are polynomially ergodic of order $m
  >1$. Further, assume that the stage 2 Markov chains
  $\Phi_1, \ldots, \Phi_k$ are polynomially ergodic of order $m$, and
  for some $\delta >0$
  $E_{q_l} |u^\pi(X;\bd_{\bq})|^{2+\delta} < \infty$ and $E_{q_l} |v^{[f],\pi}(X;\bd_{\bq})|^{2+\delta} < \infty$ for each
  $\pi \in \Pi$ and $l = 1,\cdots,k$ where
  $m > 1+2/\delta$. Then as $n_1, \ldots,n_k \rightarrow \infty$,
  \begin{equation}
    \label{eq:mclteta}
  \sqrt{n} \left(\begin{array}{c}
    \hatbd_{\bq} - {\bd}_{\bq}\\
    \hat{\bbeta}^{[f]}(\bpi; \hatbd_{\bq}) - {\bf E}_{\bpi} f\\
\end{array} \right)
\cd N\Bigg(0, \left(\begin{array}{cc}
    \varpi V_{\bq} & \Delta_{12}\\
    \Delta_{21} & \Delta_{22}\\
\end{array} \right)\Bigg),
  \end{equation}
where $\Sigma_{21} = \varpi E(\bpi; \bd_{\bq}) V_{\bq}$,
$\Sigma_{12} = \Sigma_{21}^{\top}$, and $\Sigma_{22} = \varpi E(\bpi; \bd_{\bq}) V_{\bq} E(\bpi; \bd_{\bq})^{\top} + {\boldsymbol \rho} (\bd_{\bq})$.
\item[(b)]  Suppose that the conditions of Theorem 1
  hold for the stage 1 Markov chains. Let $\widehat{V}_{\bq}$ be the consistent estimator of
  $V_{\bq}$ given in Theorem 1.  Assume that the Markov
  chains $\Phi_1, \ldots, \Phi_k$ are polynomially ergodic of order
  $m \ge (1+\epsilon)(1 + 2/\delta)$ for some $\epsilon, \delta >0$
  such that $E_{q_l} \|{\bf u}^{\bpi}(X ;\bd)\|^{4+\delta} < \infty$ and $E_{q_l} \|{\bf v}^{[f], \bpi}(X ;\bd)\|^{4+\delta} < \infty$,
  ($\|\cdot\|$ denotes the Euclidean norm) for all $l= 1,\ldots,k$, and
  $w_{n_l}$ and $b_{n_l}$ satisfy conditions 1-4 in \cite[][Theorem
  2]{vats:fleg:jone:2015}. Then
  $(n/N) \widehat{E}(\bpi; \hatbd_{\bq})
  \widehat{V}_{\bq} \widehat{E}(\bpi;
  \hatbd_{\bq})^{\top} + \boldsymbol{\hat{\rho}}(\hatbd_{\bq})$ is a strongly consistent
  estimator of $\Delta_{22}$ and
  $(n/N) \widehat{E}(\bpi; \hatbd_{\bq})
  \widehat{V}_{\bq}$ is a consistent estimator of $\Delta_{21}$.
  \end{enumerate}}

Using similar arguments as in Section~3.3 of the paper, the joint
entropy of
$\hat{\bbeta}^{[f]}(\bpi; \hatbd_{\bq})$
and $\hatbd_{\bq}$ is sum of the entropy of $\hatbd_{\bq}$, and
the conditional entropy of
$\hat{\bbeta}^{[f]}(\bpi; \hatbd_{\bq})$
given $\hatbd_{\bq}$. Thus the maximum entropy selection of skeleton
points boils down to choosing $\bq$ by maximizing $\log \mbox{det}(\widehat{V_{\bq}})$.

\begin{proof}[Proof of Theorem 4]
Since the Markov chains used in stage 1 are polynomially ergodic of
order $m > 1$, from \citet[][Theorem 1]{roy:tan:fleg:2018}, we have
$N^{1/2} (\hatbd_{\bq} - \bd_{\bq}) \cd {\cal N} (0,
V_{\bq})$. Since $n/N \rightarrow \varpi$, it follows that
$\sqrt{n} (\hatbd_{\bq} - \bd_{\bq}) \cd {\cal N} (0, \varpi
V_{\bq})$. Following \citet[][Proof of Theorem 3]{roy:tan:fleg:2018}
we write
\begin{equation}
  \label{eq:hvminv}
  \sqrt{n}(\hat{\bbeta}^{[f]}(\bpi; \hatbd_{\bq}) - {\bf E}_{\bpi} f) = \sqrt{n}(\hat{\bbeta}^{[f]}(\bpi; \hatbd_{\bq}) - \hat{\bbeta}^{[f]}(\bpi; \bd_{\bq})) + \sqrt{n}(\hat{\bbeta}^{[f]}(\bpi; \bd_{\bq}) - {\bf E}_{\bpi} f).
\end{equation}
The 2nd term involves randomness only from the
2nd stage Markov chains. Note that
\begin{equation*}
    \hat{\bf v}(\bpi; \bd_{\bq}) \cas \sum_{l=1}^k a_l E_{\pi_{\xi_l}} {\bf v}^{[f], \bpi}(X; \bd_{\bq}) = {\bf E}_{\bpi} f \odot {\bf u}(\bpi, q_1).
\end{equation*}
Since $\sum_{l=1}^k a_l E_{q_l}
{\bf u}^{\bpi}(X; \bd_{\bq}) = {\bf u}(\bpi, q_1)$,
we have
\begin{equation}
  \label{eq:jtclt}
  \sqrt{n} \left(\begin{array}{c}
    \hat{\bf v}(\bpi; \bd_{\bq}) - {\bf E}_{\bpi} f \odot {\bf u}(\bpi, q_1)\\
   \hat{\bf u}(\bpi;
   \bd_{\bq}) - {\bf u}(\bpi,
   q_1)\\
\end{array}
\right) =
  \sum_{l=1}^k a_l \sqrt{\frac{n}{n_l}} \frac{1}{\sqrt{n_l}} \sum_{i=1}^{n_l} \left(\begin{array}{c}
     {\bf v}^{[f], \bpi}(X_i^{(l)}; \bd_{\bq}) - E_{q_l}
{\bf v}^{[f], \bpi}(X; \bd_{\bq}) \\
   {\bf u}^{\bpi}(X_i^{(l)}; \bd_{\bq}) - E_{q_l} {\bf u}^{\bpi}(X;
     \bd_{\bq})\\
\end{array}
\right) .
\end{equation}
Since $\Phi_l$ is polynomially ergodic of order $m$ and
$E_{q_l} |u^{\pi}(X; \bd_{\bq})|^{2+\delta}$ and
$E_{q_l} |v^{[f],\pi}(X;\bd_{\bq})|^{2+\delta} < \infty$ are finite
for each $\pi \in \Pi$ where
$m > 1 + 2/\delta$, it follows that
\[
\frac{1}{\sqrt{n_l}} \sum_{i=1}^{n_l} \left(\begin{array}{c}
     {\bf v}^{[f], \bpi}(X_i^{(l)}; \bd_{\bq}) - E_{q_l}
{\bf v}^{[f], \bpi}(X; \bd_{\bq}) \\
   {\bf u}^{\bpi}(X_i^{(l)}; \bd_{\bq}) - E_{q_l} {\bf u}^{\bpi}(X;
     \bd_{\bq})\\
\end{array}
\right)  \cd N(0, \Lambda_l(\bd_{\bq}))
\]
 where
$\Lambda_l(\bd_{\bq})$ is defined in \eqref{eq:lamd}.
As $n_l/n \rightarrow s_l$ and the Markov chains $\Phi_l$'s are
independent, it follows that
\[
  \sqrt{n} \left(\begin{array}{c}
    \hat{\bf v}(\bpi; \bd_{\bq}) - {\bf E}_{\bpi} f \odot {\bf u}(\bpi, q_1)\\
   \hat{\bf u}(\bpi, q_1;
   \bd_{\bq}) - {\bf u}(\bpi,
   q_1)\\
\end{array}
\right) \cd N(0, \Lambda(\bd_{\bq}))).
\]
Then applying the delta method to the function ${\bf h}$ we have
a CLT for the estimator $\hat{\bbeta}^{[f]}(\bpi; \bd_{\bq})$, that
is, we have
$\sqrt{n}(\hat{\bbeta}^{[f]}(\bpi; \bd_{\bq}) - {\bf E}_{\bpi} f) \cd N(0,\boldsymbol{\rho}(\bd_{\bq}))$.

Next by Taylor series
expansion of $L(\bd) = \hat{\eta}^{[f]}(\pi; \bd)$ about $\bd_{\bq}$, we have
\begin{equation*}
  \label{eq:Ltayl}
  \sqrt{n}(L(\hatbd_{\bq}) - L(\bd_{\bq})) = \sqrt{n} \nabla L(\bd_{\bq})^{\top} (\hatbd_{\bq} - \bd_{\bq}) + \frac{\sqrt{n}}{2} (\hatbd_{\bq} - \bd_{\bq})^{\top} \nabla^2 L(\bd^*) (\hatbd_{\bq} - \bd_{\bq}),
\end{equation*}
where $\bd^*$ is between $\bd_{\bq}$ and $\hatbd_{\bq}$. As in
\cite{roy:tan:fleg:2018}, we can then show that
\[
\sqrt{n}(\hat{\eta}^{[f]}(\pi; \hatbd_{\bq}) - \hat{\eta}^{[f]}(\pi; \bd_{\bq})) = \sqrt{\varpi} e(\pi_{\xi}; \bd_{\bq}) \sqrt{N} (\hatbd_{\bq} - \bd_{\bq}) + o_p(1) .
\]
Accumulating the terms for all $\pi \in \Pi$, we have
\[
\sqrt{n}(\hat{\bbeta}^{[f]}(\bpi; \hatbd_{\bq}) - \hat{\bbeta}^{[f]}(\bpi; \bd_{\bq})) = \sqrt{\varpi} E(\bpi; \bd_{\bq}) \sqrt{N} (\hatbd_{\bq} - \bd_{\bq}) + {\bf o}_p(1) .
\]
Thus for constant vectors $t_1$ and $t_2$ of dimensions $k-1$ and
$p$ respectively, we have
\begin{align}
\label{eq:crwol}
 & t_1^{\top} \sqrt{n} (\hatbd_{\bq} - \bd_{\bq}) + t_2^{\top} \sqrt{n} (\hat{\bbeta}^{[f]}(\bpi; \hatbd_{\bq}) - \hat{\bbeta}^{[f]}(\bpi; \bd_{\bq})) \nonumber\\
&\cd N(0, \varpi(t_1^{\top} + t_2^{\top}E(\bpi; \bd_{\bq}))V_{\bq} (t_1 +E(\bpi; \bd_{\bq})^{\top}t_2) + t_2^{\top} {\boldsymbol \rho} (\bd_{\bq})t_2),
\end{align}
where the last step follows from the independence of the Markov chains
involved in the two stages. Note that the variance in \eqref{eq:crwol}
is the same as
\[ (t_1^{\top}, t_2^{\top})\left(\begin{array}{cc}
                                                 \varpi V_{\bq} & \Delta_{12}\\
                                                 \Delta_{21} & \Delta_{22}\\
\end{array} \right) (t_1^{\top}, t_2^{\top})^{\top}.
\]
Hence the Cram\'{e}r-Wold device implies the joint CLT in
\eqref{eq:mclteta}. Thus Theorem~4~(a) is proved.

From the proofs of Theorem 1 and Theorem 2~(b), we know that
$\varpi \widehat{E}(\bpi; \hatbd_{\bq})
\widehat{V}_{\bq} \widehat{E}(\bpi;
\hatbd_{\bq})^{\top}$ is a consistent estimator of
$\varpi E(\bpi; \bd_{\bq}) V_{\bq}
E(\bpi; \bd_{\bq})^{\top}$. Also,
$\nabla {\bf h}(\hat{\bf v}^{[f]}(\bpi;
\hatbd_{\bq}), \hat{\bf u}(\bpi; \hatbd_{\bq})) \cas \nabla {\bf h}({\bf
  E}_{\bpi} f \odot {\bf
  u}(\bpi, q_1), {\bf
  u}(\bpi, q_1))$.  If $\bd_{\bq}$
is known from the assumptions of Theorem~4~(b) and the results in
\citet{vats:fleg:jone:2015}, we know that $\Lambda_l (\bd_{\bq})$ is
consistently estimated by its SV estimator
$\widehat{\Lambda}_l (\bd_{\bq})$ defined in \eqref{eq:lamd}. Then
using similar arguments as in the proof of Theorem 3, we can show that
every element of
$\widehat{\Lambda}_l (\bd_{\bq})- \widehat{\Lambda}_l (\hatbd_{\bq})$
converges to zero (a.e.). Thus
$ \boldsymbol{\hat{\rho}}(\hatbd_{\bq})$ is a consistent estimator of
$ \boldsymbol{\rho}(\bd_{\bq})$.  Hence Theorem~5~(b) is proved.
\end{proof}

\section{Algorithms for computing the optimal skeleton set}
\label{sec:algorithms-suppl}

In many cases, searching over the whole space $Q$ to find the optimal set
of proposal densities is computationally hard. Often, $Q$ comprises of a
parametric family of densities parameterized by $\xi \in \Xi$ so the problem becomes choosing the skeleton set $\bxi = \{\xi_1,\xi_2,\ldots,\xi_k\}$
corresponding to the parameter values of the proposal densities that
minimizes an optimality criterion $\phi(\bxi)$.

For
convenience we work with a discretized version $\tilde\Xi$ of ${\Xi}$ and
the skeleton set $\bxi$ is constrained to be $\bxi \subset \tilde\Xi$. Finding
the optimal skeleton set in this context has been studied in the sampling
design and computer experiments literature
\citep{ko:lee:quey:1995,royl:nych:1998,fang06:comput_exper}. Here we
present the two algorithms we used in this paper, a point-swapping
algorithm and a simulated annealing algorithm. Further details for these
algorithms can be found in \cite{royl:nych:1998} and \cite{beli:1992}
respectively. 

\subsection{Point-swapping algorithm}
\label{sec:point-swapp-algor}

This algorithm performs many iterations, so it is mostly suited in cases
where the optimality criterion is fast to compute. We used the
point-swapping algorithm for computing the space-filling proposal
distributions because in these cases the criterion depends only on pairwise
distances of the proposal distributions which are fast to compute.

The algorithm proceeds as follows:
\begin{enumerate}
\item[]\textbf{Initialization:} Initialize the skeleton set at $\bxi^{(0)}$ with $|\bxi^{(0)}| =k$.
\item[]\textbf{Iterations:} Repeat for $i=1,2,\ldots$:

  For $j=1,\ldots,k$: Swap the $j$th element of $\bxi^{(i-1)}$ with
  the element $\xi_j' \in \tilde\Xi\setminus\bxi^{(i-1)}$ such that the new
  set $\bxi^{(i)} = \bxi^{(i-1)} \setminus \{\xi_j\} \cup \{\xi_j'\}$
  produces the biggest drop in the value of the optimality criterion. If
  no such $\xi_j'$ exists, i.e., if $\phi(\bxi^{(i-1)}) < \phi(\bxi^{(i-1)}
  \setminus \{\xi_j\} \cup \{\xi'\})$ for all $\xi' \in \tilde\Xi\setminus\bxi^{(i-1)}$, then $\bxi^{(i)} = \bxi^{(i-1)}$.

\item[]\textbf{Termination:} Stop if, after looping over all elements in
  $\bxi^{(i)}$, the skeleton set remains unchanged. Return the final
  skeleton set.
\end{enumerate}

We used the implementation in the R package \texttt{fields} \citep{R:fiel}
to compute the optimal skeleton set in our paper.

\subsection{Simulated annealing algorithm}
\label{sec:simul-anne-algor}

We use simulated annealing to compute the optimal set of proposal
distributions for MNX and ENT. The criterion for these methods is based on
the SV estimate of the asymptotic SE of the multiple IS estimator which is
computed from Monte Carlo samples. In this case, if samples from a
particular proposal distribution in the skeleton set already exist, then
they are reused, otherwise they are generated and stored for a possible
future use.

The algorithm proceeds as follows:
\begin{enumerate}
\item[]\textbf{Initialization:} Initialize the skeleton set at
  $\bxi^{(0)}$ with $|\bxi^{(0)}| =k$ and an initial temperature at $T_0$.
\item[]\textbf{Iterations:} Repeat for $i=1,2,\ldots$:
  \begin{enumerate}
  \item Set $T = T_0/\log(\lfloor(i-1)/B\rfloor B + \exp(1))$, where $B$ is a
    parameter of the algorithm denoting the number of iterations before the
    temperature is lowered.
  \item Randomly select $\xi \in \bxi^{(i-1)}$ and $\xi' \in \tilde\Xi \setminus
    \bxi^{(i-1)}$. Form the candidate set $\bxi' = \bxi^{(i-1)} \setminus \{\xi\}
    \cup \{\xi'\}$.
  \item With probability
    $\min\{1, \exp[(\phi(\bxi^{(i-1)}) - \phi(\bxi'))/T]\}$ set
    $\bxi^{(i)}=\bxi'$, otherwise set $\bxi^{(i)}=\bxi^{(i-1)}$.
  \end{enumerate}
\item[]\textbf{Termination:} Stop if $i > i_\mathrm{max}$, a predetermined
  number of iterations. Return the skeleton set found among all iterations
  which corresponds to the lowest value of the optimality criterion.
\end{enumerate}

\section{Finney's (1947) vasoconstriction data analysis using robit model}
\label{sec:robit}


\pcite{finn:1947} vasoconstriction data consist of 39 binary responses
denoting the presence or absence, \(y\), of vasoconstriction on the
subject's skin after he or she inhaled air of volume \(V\) at rate
\(R\). We consider a binomial generalized linear model (GLM) where the probability of presence for the
\(i\)th subject, \(\alpha_i\), is modeled using a robit link function
with degrees of freedom (df) \(\xi\),
$F_\xi^{-1}(\alpha_i) = \beta_0 + \beta_1\log V_i + \beta_2\log R_i $,
for \(i=1,\ldots,39\). Here, $F_\xi (\cdot)$ denotes the distribution
function of the standard Student's $t$ distribution with df \(\xi\).
As in \cite{roy:2014}, we consider a Bayesian analysis of the data
with robit model. The prior for $\beta$ is
$\beta \sim t_3(0, 10^{4} (W^\top W)^{-1}, 3)$, where
\(W\) is the design matrix.

\cite{roy:2014} estimates the df parameter \(\xi\) by maximizing the
marginal likelihood, that is, $\hat{\xi} = \argmax \theta_\xi(\by)$. 
In particular, \cite{roy:2014} uses the multiple IS estimator
(1.2) to estimate the (ratios of) marginal likelihoods,
which in turn provides the estimate $\hat{\xi}$. Our objective is to
choose the importance sampling distributions from the family of
posterior densities $\Pi = \{\pi_\xi(\beta|\by): \xi > 0\}$ for the
estimation of (1.2). We consider this
in section~\ref{sec:gisrob}. Whereas in
section~\ref{sec:vasoc-MVN}, we analyze this problem with proposal
densities from the multivariate normal family.

\subsection{Selection of proposals for multiple IS}
\label{sec:gisrob}
In this section, we consider $Q=\Pi$, thus, choosing proposal
distributions is the same as choosing appropriate $\xi$
values. Because $\xi$ represents the df parameter, we
consider a wide range of points $\tilde\Xi=\{0.1, 0.2, \ldots{}, 20\}$ from
where we choose the skeleton sets for the different methods. The reference
density corresponds to $\tilde\xi$, where $\tilde\xi = 10$ is at the middle
of this range. For the multiple IS methods we choose $k=5$ points, one of which must be
$\tilde\xi$.

The computation is done in two phases. In the
first phase we find the optimal skeleton set, $\bxi$, for each method. In the second
phase we compare the relative standard errors of the naive and multiple IS estimators
using the skeleton sets computed in the first phase. The total number of
samples used for each method in the second phase is kept the same. The
required Markov chain samples were generated by Hamiltonian Monte Carlo
implemented in the stan language \citep{stan:2017}. In all
calculations involving the asymptotic variance of the IS estimator, the SV
estimate with the Tukey-Hanning window,
\[
  w_{n} (j) = 0.5[1 + \cos (\pi |j|/ b_{n})]I(|j| < b_{n}),
\]
was used, where $b_n = \sqrt{n}$.

\noindent{\bf Phase 1: Finding the optimal set of proposal densities}

\begin{figure}
  \centering
  \includegraphics[width=.5\linewidth]{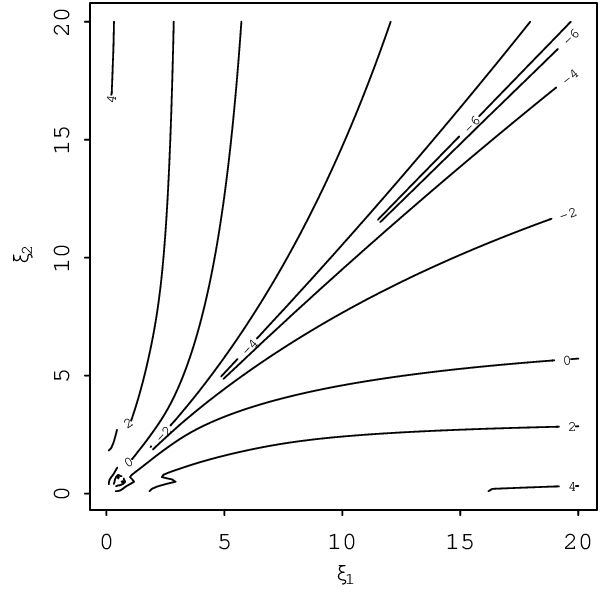}
  \caption{ Contour plot of the logarithm of the SKLD between the
     densities corresponding to $\xi_1$ and $\xi_2$ for the vasoconstriction
     example. }
  \label{fig:vasoc_skld}
\end{figure}

\begin{itemize}
\item \textbf{NIS:} This is not required because the proposal density
  in the naive importance sampling (NIS) is fixed at
  $\bxi_\mathrm{nis} = \{10\}$.
\item \textbf{SFE:} This method is based on the Euclidean distance
  between the parameters. Therefore,
  $\bxi_\mathrm{sfe} = 
  \{2, 6, 10, 14, 18\}$.
\item \textbf{SFS:} This method requires the SKLD between two
  densities corresponding to two $\xi$ values. Figure~\ref{fig:vasoc_skld} shows the logarithm of pairwise
  SKLD between densities corresponding to two different values of
  $\xi$. The SKLD is computed using the approximation of
  Section~\ref{sec:laplace}. It can be seen that the distance is
  non-Euclidean. For example, the densities corresponding to $\xi_1=1$
  and $\xi_2=5$ are further apart than the densities corresponding to
  $\xi_1=16$ and $\xi_2=20$. Using the algorithm of
  Section~\ref{sec:point-swapp-algor} we select
  $\bxi_\mathrm{sfs} = \{ 0.3, 1.1, 1.9, 3.3, 10 \}$.  It is noted
  that the points concentrate more on the low values in $\tilde\Xi$.
\item \textbf{SEQ:} In this case the optimal set is computed by starting at
  $\bxi^{(1)} = \bxi_\mathrm{nis}$. Then, given that at the $i$th
  iteration, $i=2,\ldots,k$ we are at $\bxi^{(i-1)}$, we obtain $\bxi^{(i)}
  = \bxi^{(i-1)} \cup \{\xi'\}$, where $\xi'$ corresponds to the point in
  $\tilde\Xi\setminus\bxi^{(i-1)}$ with the highest relative standard error. The relative standard error
  is again computed using 2,000  samples for stage~1 and 2,000
  new samples for stage~2, after a burn in of 400. We find
  $\bxi_\mathrm{seq} = 
  \{0.1, 0.2, 0.3, 0.7, 10\}$.
\item \textbf{MNX:} The optimal set is found by simulated annealing
  (Section~\ref{sec:simul-anne-algor}). We start the simulated annealing
  algorithm at $\bxi_\mathrm{sfs}$, and perform $i_\mathrm{max}=250$
  iterations with $T_0=0.1$ and $B=10$. The optimality criterion is
  computed as follows. Using 2,000 Markov chain samples for stage~1 and
  2,000 samples for stage~2, we compute $\hat\upsilon_1^2(\xi) := \hat{c}(\pi;\hat{\bd})^{\top} \hat{V}
\hat{c}(\pi;\hat{\bd})$ and
$\hat\upsilon_2^2(\xi) := \hat{\tau}^2(\pi;\hat{\bd})$ given in
Theorem~2(a) for each $\xi \in \tilde\Xi$ as well as
$\hat{u}(\xi) = \hat{u}$ given in equation~(1.2) of the main
paper. Then, given stage~1 sample size of $N$ and stage~2 sample
size of $n$, the relative standard error estimate is given by
\begin{equation}
  \label{eq:relse}
  \mathrm{RelSE}(\xi,N,n) :=\frac{\hat\upsilon_1(\xi)/\sqrt{N} +
    \hat\upsilon_2(\xi)/\sqrt{n}}{\hat u(\xi)}.
\end{equation}
Assuming that the total sample size $M=N+n$ is fixed, the objective becomes
choosing $\bxi_\mathrm{mnx}$ in order to minimize
\begin{equation}
  \label{eq:relsen}
  \min_{N \in (0,M)} \max_{\xi\in\tilde\Xi} \mathrm{RelSE}(\xi,N,M-N).
\end{equation}
We find $\bxi_\mathrm{mnx} = \{0.1, 0.4, 1.6, 3.3, 10\}$. One can
also impose a constraint that $N$ is at least some
number and at most some other number while finding $\bxi_\mathrm{mnx}$ minimizing \eqref{eq:relsen}.

\item \textbf{ENT:} The optimal set is found by simulated annealing
  (Section~\ref{sec:simul-anne-algor}). We start the simulated annealing
  algorithm at $\bxi_\mathrm{sfs}$, and perform $i_\mathrm{max}=250$
  iterations with $T_0=1$ and $B=10$. The optimality criterion used for
  this method is $-\log\mbox{det}(U)$ where $U$ is the matrix with $(i,j)$th
  element $U_{ij} = (\widehat{V}_{\bq})_{ij}/(\hat{d}_i\hat{d}_j)$. The
  estimates $\hat{\bd}$ and $\widehat{V}_{\bq}$ are computed from 2,000
  Markov chain samples. In this example, the ENT skeleton set turns out to
  be the same as the MNX set, thus 
  $\bxi_\mathrm{ent} = \bxi_\mathrm{mnx} = \{0.1, 0.4, 1.6, 3.3, 10\}$.
\end{itemize}

\noindent{\bf Phase 2: Estimation of the ratio of marginal densities}

After the skeleton sets are found, we generate a total of $M=50,000$
samples from the proposal densities, equally divided among all
densities in the set. The total sample size $M$ is generally
determined by available computational resources. Since we have derived
consistent SV estimators of the asymptotic variance of $\hat{u}$, the
sample size $M$ can be chosen such that the overall SE of $\hat{u}$ is
smaller than a pre-determined level of accuracy. For NIS we simply
take $M$ Gibbs samples from the density corresponding to
$\bxi_\mathrm{nis}$, and for the multiple IS methods we take $M/k$
samples from each density in the corresponding $\bxi$ set. However,
the total of $M/k$ samples must be split into stage~1, for estimating
the ratio of marginals within $\bxi$, $\bd$, and stage~2, for
estimating the ratio of marginals over the whole set $\tilde\Xi$. To
determine the optimal split we use equation~\eqref{eq:relse} where
$\hat\upsilon_1(\xi)$ and $\hat\upsilon_2(\xi)$ are calculated from
2,000 stage~1 and 2,000 stage~2 Markov chain samples. For MNX, SEQ,
and ENT the existing samples from Phase~1 are reused but for the
space-filling methods, new samples are generated. We take an equal
number of samples from each density, thus we take $\hat N/k$ stage~1
samples from each density in the skeleton set where $\hat N$ is the
integer that minimizes
$\max_{\xi \in \tilde\Xi} \mathrm{RelSE}(\xi,N,M-N)$, and $\hat n/k$
stage~2 samples, where $\hat n = M - \hat N$. This corresponds to
stage~1 sample sizes of 1500, 6500, 8500, 9000, 9000 and stage~2
sample sizes of 8500, 3500, 1500, 1000, 1000 from each density for
SFE, SFS, SEQ, MNX, ENT, respectively.

\begin{figure}
  \includegraphics[width=.49\linewidth]{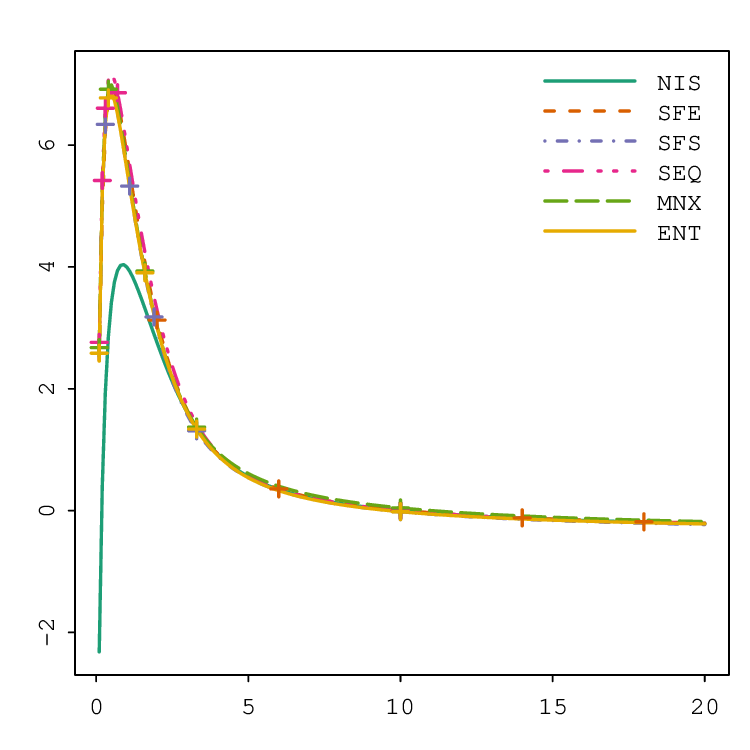}
  \includegraphics[width=.49\linewidth]{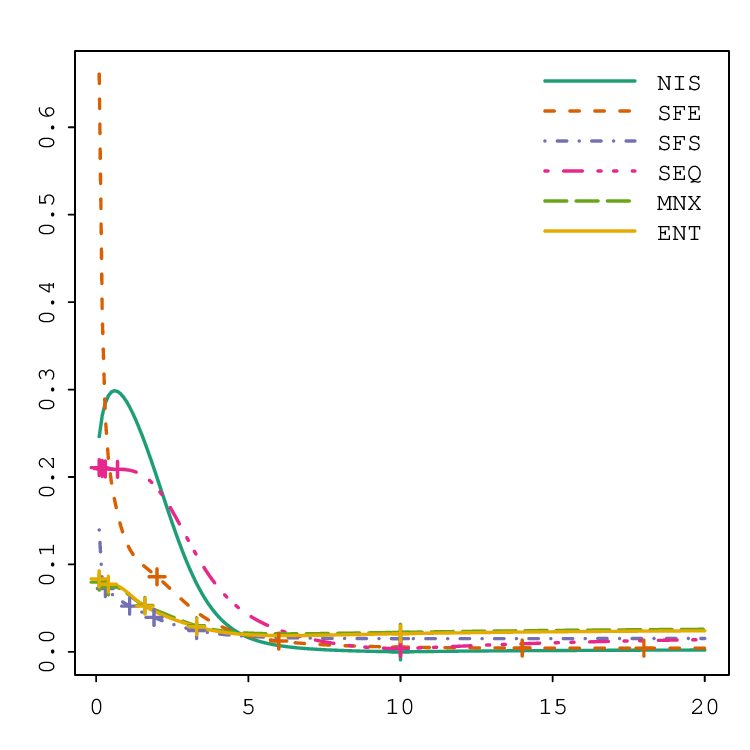}
 \caption{Vasoconstriction example: The left and right panels show the
   values of $\log \hat{u}(\hatbd_\bxi)$ and estimates of the relative standard error
   of $\hat{u}(\hatbd_\bxi)$ across $\xi$ values for different skeleton sets,
   respectively. Points included in the skeleton sets are indicated by $+$.} 
\label{fig:vasoc_SE}
\end{figure}

The estimates of the relative standard error of~(1.2) and the value of the
logarithm of~(1.2) across all \(\xi\) values in $\tilde\Xi$ corresponding to
the different skeleton sets, $\bxi$, chosen are plotted in
Figure~\ref{fig:vasoc_SE}. It can be seen that the relative standard error is
larger when $\xi$ is small. NIS has the lowest relative standard error at
$\tilde{\xi}$ but results in much higher standard errors at low values
of $\xi$. Indeed, at $\xi =0.5$, the SV estimate of the relative standard error
for NIS method is about four times larger than that for the MNX
and ENT methods. It can also be seen that SFE does not have as good
performance as SFS, again because it avoids sampling from densities
corresponding to low values of $\xi$ which are the ones that produce
the largest relative standard error. SEQ also does not have good performance
because it concentrates all points in a very narrow region while,
evidently, it is more beneficial to spread the points to cover a wider
area as MNX and ENT do.

\subsection{Multiple IS using a mixture
  of multivariate normal proposals}
\label{sec:vasoc-MVN}


The proposed methods of choosing reference distributions are
applicable to IS estimators in the situations where $Q$ can be different
from $\Pi$. To demonstrate this, in this section we analyze the
vasoconstriction data using the model and method discussed in
Section~\ref{sec:gisrob} with the difference that the proposal
densities are now chosen from the multivariate normal family instead
of the family of the posterior density of $\beta$. Thus we are now
able to draw independent and identically distributed (iid) samples from the importance sampling distributions
which is not the case when these are posterior densities of $\beta$
which require Markov chain Monte-Carlo sampling. Since $\bd$ is known, the reverse
logistic regression estimation is not needed here. The space filling, minimax and
sequential approaches developed in the main paper can be used for
selecting multivariate normal proposals as we describe below.

If $n_l$ iid samples, $X_i^{(l)}$, $i=1,\ldots,n_l$, are drawn from
the density $q_l(x)$ (a normal density described later),
$l=1,\ldots,k$, then the normalizing constant, $\theta_\xi$, of the posterior density
$\pi_\xi(x) \equiv \nu_\xi(x)/\theta_\xi$, is estimated by
\begin{equation}
  \label{eq:mvn1}
  \hat{\theta}_\xi = \frac{1}{|n|} \sum_{l=1}^k \sum_{i=1}^{n_l}
  \frac{\nu_\xi(X_i^{(l)})}{\bar q(X_i^{(l)})},
\end{equation}
where $n = n_1 + \ldots + n_k$ and $\bar q(x) = (n_1/n) q_1(x) + \ldots
+ (n_k/n) q_k(x)$. The variance of this estimator is estimated by
\begin{equation}
  \label{eq:mvn2}
  \widehat{\mbox{Var}}(\hat{\theta}_\xi) = \frac{1}{n} \sum_{l=1}^k
  \sum_{i=1}^{n_l} \left( \frac{\nu_\xi(X_i^{(l)})}{\bar q(X_i^{(l)})} -
    \hat{\theta}_\xi \right)^2.
\end{equation}

To choose the proposal densities corresponding to a skeleton set $\bxi$,
for $\xi \in \bxi$, let
$\tilde{\beta}_\xi$ denote the maximizer of $\ell_\xi(\beta|\by) \pi(\beta)$
and let $\tilde{H}_\xi$ denote the Hessian matrix of
$-\log \{\ell_\xi(\beta|\by) \pi(\beta)\}$ evaluated at
$\tilde{\beta}_\xi$. Then the normal approximation to $\pi_\xi(\beta|\by)$ is taken to be the
multivariate normal with mean $\tilde{\beta}_\xi$ and variance
$\tilde{H}^{-1}_\xi$. Thus, $q_l$ is the normal approximation to the posterior
density $\pi_{\xi}(\beta|\by)$, where $\xi=\xi_l$ is a skeleton point.

Unlike in Section~\ref{sec:gisrob}, where we used Hamiltonian
Monte Carlo to obtain approximate samples from the proposal
(posterior) densities, here we draw iid samples from the proposal
distributions. We use~\eqref{eq:mvn1} to estimate $\theta_\xi$ for all
$\xi$ values in the range $\tilde\Xi$ identified
previously.  We set $k=5$ and generate $n_l = 10000$ samples from each
of the proposal densities $q_l, l=1,\dots, 5$. We also use
naive importance sampling by drawing $5 \times 10000$ samples from the
normal approximation to the posterior for $\beta$ corresponding to
$\bxi_\mathrm{nis} = \{\tilde{\xi}=10\}$.

For the two space filling methods, the sets $\bxi_\mathrm{sfe}$ and
$\bxi_\mathrm{sfs}$ are obtained as described in
Section~\ref{sec:gisrob}, thus the skeleton sets are the same as in that
section: $\bxi_\mathrm{sfe} = \{2, 6, 10, 14, 18\}$ and $\bxi_\mathrm{sfs} = \{ 0.3, 1.1, 1.9, 3.3, 10 \}$. 

For SEQ, we aim to select the set $\bxi_\mathrm{seq}$
sequentially, starting with $\bxi^{(1)} = \{\tilde{\xi}\}$. At
the $j$th iteration, we obtain the set $\bxi^{(j)} =
\bxi^{(j-1)} \cup \{\xi_j\}$.
Then, we draw samples from the normal approximation
to the posterior density corresponding to this $\xi_j$, as described in the previous
paragraph. Using these samples, together with existing samples from $\bxi^{(j-1)}$, we compute, using~\eqref{eq:mvn1}
and~\eqref{eq:mvn2}, the relative standard error
of $\hat{\theta}_\xi$ for $\xi \in \tilde\Xi$. The value of $\xi$
corresponding 
to the highest relative standard error, denoted by $\xi_{j+1}$, is added to
the set. The final set obtained using this method is
$\bxi_\mathrm{seq} = \{0.6, 0.7, 0.8, 0.9, 10\}$.

For MNX, we run the simulated annealing algorithm with $T_0=10$, $B=10$, $i_\mathrm{max}=250$, starting from $\bxi_\mathrm{sfs}$. For every
new density added to the skeleton set we generate 4000 samples from the
corresponding normal approximation, if not already available. These
samples, along with the existing samples from the other densities
previously added are used to calculate the relative standard error,
using~\eqref{eq:mvn1} and~\eqref{eq:mvn2}, for all $\xi \in \tilde\Xi$. The
objective of the simulated annealing is to minimize the maximum relative standard error
across $\tilde\Xi$. The optimal skeleton set is
$\bxi_\mathrm{mnx} = \{0.2, 0.3, 0.7, 2.6, 10\}$.

\begin{figure}[ht]
  \centering
  \includegraphics[width=.49\linewidth]{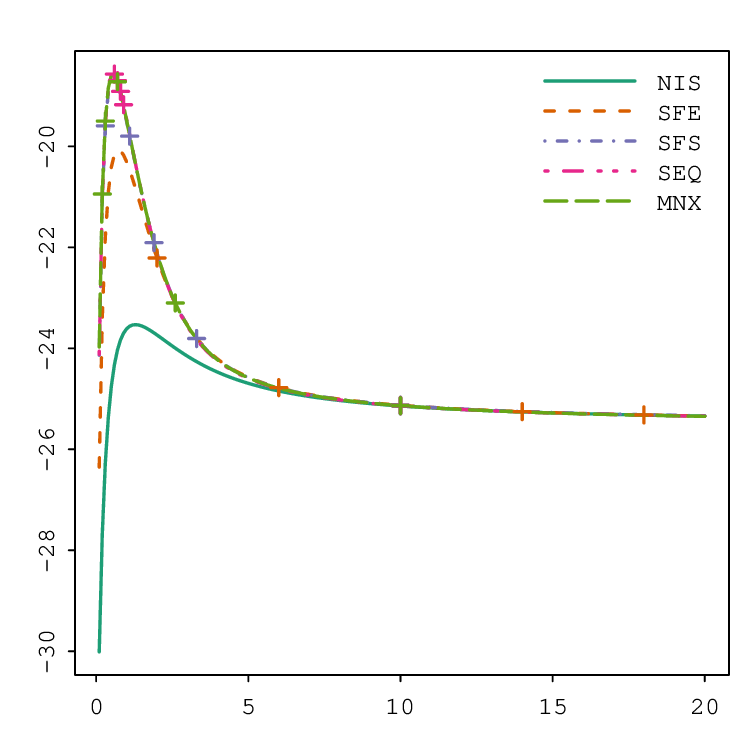}
  \includegraphics[width=.49\linewidth]{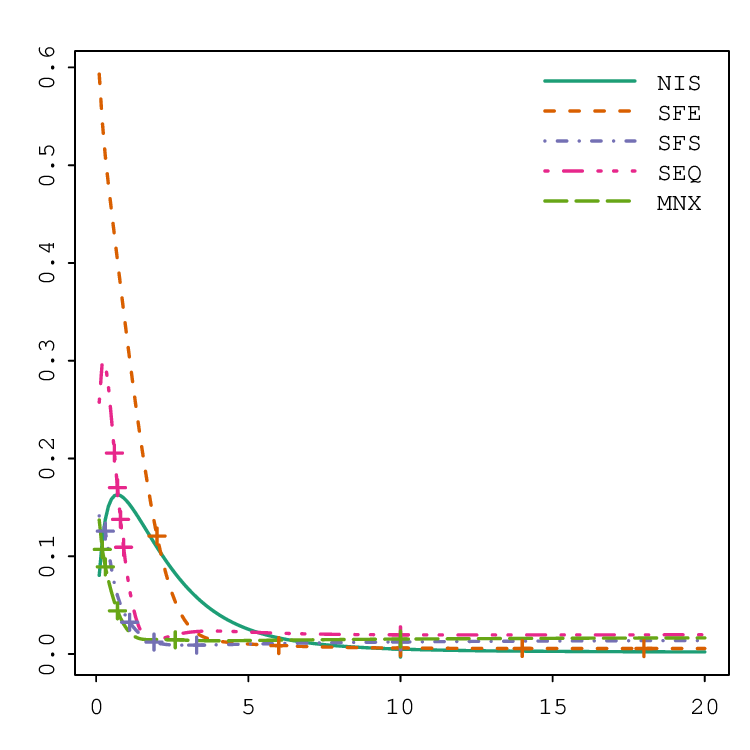}
  \caption{Logarithm of the IS estimates of the marginal density for
    Finney's vasoconstriction data using normal proposal densities
    (left); and estimates of the relative standard error of the marginal density
    estimates (right). Points included in the skeleton sets are
    indicated by $+$.}
  \label{fig:mvnmix}
\end{figure}

Plots of the estimator~\eqref{eq:mvn1} using the different methods are
shown in the left panel of
Figure~\ref{fig:mvnmix}. It can be seen that NIS and SFE are significantly
different from the other methods. 
The plots of the relative standard error estimates using~\eqref{eq:mvn2} are  shown
in the right panel of
Figure~\ref{fig:mvnmix}. It can be seen that the NIS, SFE, and SEQ methods lead
to significantly higher relative standard error compared to the proposed methods for
low values of $\xi$. For example, at $\xi=0.5$, the relative standard error for NIS,
SFE, and SEQ are
2.5, 7, and 3.8 times higher respectively than MNX.

\section{Further details on the autologistic example used in the main
  paper}
\label{sec:autologistic-suppl}

\subsection{Derivation of the autologistic pmf}
\label{sec:auto}
We start with the conditional pmf of $x(s_i) | \bx_{-s_i}$ given by
\[
  \pi_i(x(s_i) | \bx_{-s_i}) = p_i^{x(s_i)} (1- p_i)^{1- x(s_i)},
\]
where, as in the main paper,
\[
p_i = \frac{\exp\{   \mbox{logit}(\kappa)+ (\gamma/w) \sum_{s_j \in
      \mbox{nb}_i} (x(s_j) - \kappa)\}}{1 + \exp\{ \mbox{logit}(\kappa)+ (\gamma/w) \sum_{s_j \in
      \mbox{nb}_i} (x(s_j) - \kappa)\}}.
\]
Letting $\pi(\bx | \gamma, \kappa) \propto \exp(A(\bx | \theta))$,
since only pairwise dependencies are considered, it is known that \citep{kais:cres:2000}
\[
  A(\bx | \theta) = \sum_{1 \le i \le m} \log \frac{\pi_i(x(s_i) | \bx^*_{-s_i})}{\pi_i(x^*(s_i) | \bx^*_{-s_i})} +
  \sum_{1 \le i < j \le m; s_j \in \mbox{nb}_i} \log \frac{\pi_i(x(s_i) | x(s_j), \bx^*_{-\{s_i, s_j\}})\pi_i(x^*(s_i) | \bx^*_{-s_i})}{\pi_i(x^*(s_i) | x(s_j), \bx^*_{-\{s_i, s_j\}})\pi_i(x(s_i) | \bx^*_{-s_i})},
  \]
  for a suitably chosen $\bx^*$. Choosing $\bx^* = {\bf 0}$, simple calculations show that
  \[
    A(\bx | \theta) = (\mbox{logit}(\kappa) - \gamma \kappa) \sum\nolimits_{i=1}^m x(s_i)+ \frac{\gamma}{2w} \sum\nolimits_{i=1}^m \sum\nolimits_{s_j \in
    \mbox{nb}_i} x(s_i) x(s_j).
  \]

\subsection{Computational details}
\label{sec:comp-autologistic-suppl}

Here we give more details on how we produced the results for the
autologistic example in the Section~4 of the main paper. We also
present the case where $\kappa$ remains fixed at $0.5$ and only
$\gamma$ is allowed to vary. The parameter set where we search over is
\begin{equation*}
  \gamma \in \{-4, -3.9, \ldots, 4\},
\end{equation*}
in the case where $\kappa$ is set fixed at $\kappa=0.5$, and
\begin{equation*}
  (\gamma,\kappa) \in \{-4, -3.2, \ldots, 4\} \times \{0.1, 0.2, \ldots, 0.9\},
\end{equation*}
in the case where both parameters are assumed unknown. In both cases we denote the
parameters by $\xi$ and the parameter set by $\tilde\Xi$.

The computation is done in two phases. In the
first phase we find the optimal skeleton set, $\bxi$, for each method. In the second
phase we compare the relative standard error of the naive and multiple IS estimators
using the skeleton sets computed in the first phase. The total number of
samples used for each method in the second phase is kept the same. The
required Markov chain samples were generated by Gibbs sampling from the
conditional distribution of each component of $\bx$ given its neighbors. In all
calculations involving the asymptotic variance of the IS estimator, the SV
estimate with the Tukey-Hanning window,
\[
  w_{n} (j) = 0.5[1 + \cos (\pi |j|/ b_{n})]I(|j| < b_{n}),
\]
was used, where $b_n = \sqrt{n}$.

\noindent{\bf Phase 1: Finding the optimal set of proposal densities}

\begin{itemize}
\item \textbf{NIS:} This is not required because the proposal density is
  fixed at $\bxi_\mathrm{nis} = \{0\}$ and $\bxi_\mathrm{nis} = \{(0,0.5)\}$ for the cases
  $\kappa$ fixed and $\kappa$ varying, respectively.
\item \textbf{SFE:} This method is based on the Euclidean distance
  between the parameters. Each component of the parameter set $\tilde\Xi$ is
  scaled to vary between 0 and 1 before the optimal set is
  computed. We find
  $\bxi_\mathrm{sfe} = \{-3.2, -1.6, 0.0, 1.6, 3.2\}$ and
  $\bxi_\mathrm{sfe} =
  \{(0,0.5),(-2.4,0.3),(-2.4,0.7),\allowbreak (2.4,0.3),(2.4,0.7)\}$ for the cases $\kappa$
  fixed and $\kappa$ varying, respectively.
\item \textbf{SFS:} This method requires the SKLD between two pairs
  $\xi_1$, $\xi_2$. We compute the integrals in~(3.7a) of the main
  paper by Monte Carlo using 3,000 Gibbs samples after a burn-in of
  400 from each distribution for the $\kappa$ known case, and 20,000
  samples after a burn-in of 4,000 for the $\kappa$ estimated case. The
  point-swapping algorithm of Section~\ref{sec:point-swapp-algor} is
  used to find the optimal set. We find
  $\bxi_\mathrm{sfs} = \{-3.36, -1.68, 0.00, 1.68, 3.28\}$ and
  $\bxi_\mathrm{sfs} = \{(0,0.5),(-3.2,0.5),\allowbreak
  (-1.6,0.8),(-1.6,0.2),(3.2,0.5)\}$ for the cases $\kappa$ fixed and
  $\kappa$ varying, respectively.
\item \textbf{MNX:} The optimal set is found by simulated annealing
  (Section~\ref{sec:simul-anne-algor}). We start the simulated annealing
  algorithm at $\bxi_\mathrm{sfs}$ and perform $i_\mathrm{max}=250$
  iterations with $T_0=10$ and $B=10$. 
  The SV
  estimates were calculated using 3,000 Gibbs samples for stage~1 and 3,000
  new samples for stage~2 for the $\kappa$ known case, and 20,000
  samples for stage~1 and 20,000 samples for stage~2 for the $\kappa$ estimated case. We find
  $\bxi_\mathrm{mnx} = \{-3.20, -1.60, 0.00, 1.68, 3.20\}$ and
  $\bxi_\mathrm{mnx} =
  \{(0,0.5),(-3.2,0.5),(-1.6,0.2),(-0.8,0.8),(4.0,0.4)\}$ for the cases $\kappa$ fixed
  and $\kappa$ varying, respectively.
\item \textbf{SEQ:} In this case the optimal set is computed by starting at
  $\bxi^{(1)} = \bxi_\mathrm{nis}$. Then, given that at the $i$th
  iteration, $i=2,\ldots,k$ we are at $\bxi^{(i-1)}$, we obtain $\bxi^{(i)}
  = \bxi^{(i-1)} \cup \{\xi'\}$, where $\xi'$ corresponds to the point in
  $\tilde\Xi\setminus\bxi^{(i-1)}$ with the highest relative standard error. The relative standard error
  is again computed using 3,000 Gibbs samples for stage~1 and 3,000
  new samples for stage~2 for the $\kappa$ known case, and 20,000
  samples for stage~1 and 20,000 samples for stage~2 for the $\kappa$ estimated case. We find
  $\bxi_\mathrm{seq} = \{-4.00, -3.92, 0.00, 3.92, 4.00\}$ and
  $\bxi_\mathrm{seq} =
  \{(0,0.5),(-4.0,0.8),(-2.4,0.9),(-0.8,0.1),(3.2,0.2)\}$ for the cases $\kappa$ fixed
  and $\kappa$ varying respectively.
  
\item \textbf{ENT:} The optimal set is found by simulated annealing
  (Section~\ref{sec:simul-anne-algor}). We start the simulated annealing
  algorithm at $\bxi_\mathrm{sfs}$ and perform $i_\mathrm{max}=250$
  iterations with $T_0=10$ and $B=10$. The optimality criterion used for
  this method is $-\log\mbox{det}(U)$ where $U$ is the matrix with $(i,j)$th
  element $U_{ij} = (\widehat{V}_{\bq})_{ij}/(\hat{d}_i\hat{d}_j)$. The
  estimates $\hat{\bd}$ and $\widehat{V}_{\bq}$ are computed from 3,000
  Gibbs samples for the $\kappa$ known case, and 20,000
  samples for the $\kappa$ estimated case. We find
  $\bxi_\mathrm{ent} = \{-4, -2.08, 0.00, 2.32, 3.52\}$ and
  $\bxi_\mathrm{ent} =
  \{(0,0.5),(-4.0,0.3),(-1.6,0.9),\allowbreak (-0.8,0.1),(4.0,0.4)\}$ for the cases $\kappa$ fixed
  and $\kappa$ varying respectively.
\end{itemize}


\noindent{\bf Phase 2: Estimation of the ratio of marginal densities}
  
For the case where only $\gamma$ varies, after the skeleton sets are found, we generate a total of $M=100,000$
samples from the proposal densities, equally divided among all
densities in the set. Thus, for NIS we simply take $M$ Gibbs samples
from the density corresponding to $\bxi_\mathrm{nis}$ and for the
multiple IS methods we take $M/k$ samples from each density in the
corresponding $\bxi$ set. However, the total of $M/k$ samples must be
split into stage~1, for estimating the ratio of marginals within
$\bxi$, $\bd$, and stage~2, for estimating the ratio of marginals over
the whole set $\tilde\Xi$. To determine the optimal split we use
equation~\eqref{eq:relse} where $\hat\upsilon_1(\xi)$ and
$\hat\upsilon_2(\xi)$ are calculated from 3,000 stage~1 and 3,000
stage~2 Gibbs samples. For MNX, SEQ, and ENT the existing samples from
Phase~1 are reused but for the space-filling methods, new samples are
generated. We take $\hat N/k$ stage~1 samples from each density in the
skeleton set where $\hat N$ is the integer that minimizes
$\max_{\xi \in \tilde\Xi} \mathrm{RelSE}(\xi,N,M-N)$, and $\hat n/k$
stage~2 samples, where $\hat n = M - \hat N$. In the case where
$\kappa$ is fixed, this corresponds to stage~1 sample sizes of 11000,
13000, 12000, 18000, 15000 and stage~2 sample sizes of 9000, 7000,
8000, 2000, 5000 from each density for SFE, SFS, MNX, SEQ, ENT,
respectively. In the case where $\kappa$ also varies, the stage~1
sample sizes were 5000, 11667, 13889, 25000, 31667, and the stage~2
sample sizes were 45000, 38333, 36111, 25000, 18333 for SFE, SFS, MNX,
SEQ, ENT, respectively.

\begin{figure}
  \centering
  \includegraphics[width=\linewidth]{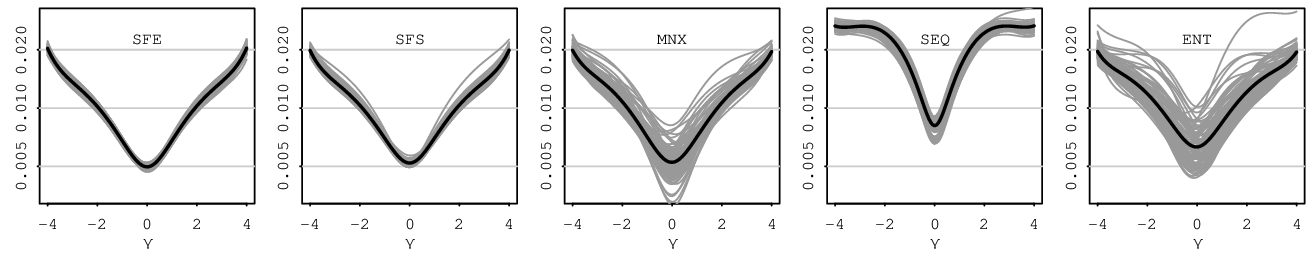}
  \caption{Relative standard error plots based on 100 replications of the autologistic model with $\kappa=0.5$
fixed and $\gamma$ varying. The bold line provides the average of the relative standard errors.}
  \label{fig:auto_repl}
\end{figure}

Finally, we study the performance of the different methods over
repeated simulations. Figure~\ref{fig:auto_repl} provides relative
standard error plots based on 100 replications of the autologistic
model with $\kappa=0.5$ fixed and $\gamma$ varying. Each time a
skeleton set is chosen and the standard error is computed based on the
Monte Carlo samples. (Naive IS suffers from high variability when
$\gamma$ is away from the origin and is not considered in
Figure~\ref{fig:auto_repl}.) From the plots we see that the SF methods
are the least variable. SFE does not require samples from the
autologistic models to choose the skeleton points. On the other hand,
SFS requires samples to compute the SKLD but even then it does not
increase the variability compared to SFE. MNX and ENT seem to be the
most variable, but shape of the relative standard error curve mostly
remains the same. The SEQ method consistently resulted in the highest
standard errors. Indeed, the maximum and the minimum of the ratios of the (average)
relative standard errors of SEQ to that of MNX are 2.4 and 1.3, respectively.

\section{Analysis of  radionuclide concentrations using spatial GLMM}
\label{rongelap-suppl}
The dataset consists of spatial measurements of \(\gamma\)-ray counts
\(y_i\) observed during \(\ell_i\) seconds at the $i$th coordinate on the
Rongelap island, \(i=1,\ldots,157\). These data were analyzed
by \cite{digg:tawn:moye:1998} and \cite{evan:roy:2019} among others using a
spatial generalized linear mixed model (SGLMM). We consider a Poisson SGLMM using
a parametric link function for the \(\gamma\)-ray counts, that is, we assume $y_i|\mu_i \ind \mathrm{Po}(\ell_i \mu_i)$ with $g_\lambda(\mu_i) = z_i$ for $i=1,\dots,157,$
where $g_\lambda(\cdot)$ is a modified Box-Cox link given in \cite{evan:roy:2019} with parameter
$\lambda$, and $z_i$'s are the latent variables.  
Let $\by$ and $\bmu$ denote the vectors of $y_i$'s and $\mu_i$'s,
respectively. Then $\bz= (z_1,\dots, z_{157})$ is modeled by a
multivariate Gaussian distribution corresponding to a Gaussian random
field (GRF) $\mathcal{Z}$ at the sampled locations. In particular, we
assume
$\mathcal{Z}|\beta,\sigma^2 \sim
\mathrm{GRF}(\beta,\sigma^2,\phi,\omega,\kappa)$, the GRF with
constant mean $\beta$, Mat\'ern correlation, variance $\sigma^2$,
range $\phi$, relative nugget $\omega$ and smoothness $\kappa$. The
partial sill parameter \(\sigma^2\) is assigned a
scaled-inverse-chi-square prior ($\mathrm{ScInv}\mathcal{X}^2(1,1)$),
and conditioned on $\sigma^2$, the mean parameter
\(\beta \sim N(0,100\sigma^2)\). Let
\(\xi = (\lambda,\phi,\omega,\kappa)\).  We consider estimating the
marginal likelihood for $\xi$ (relative to an arbitrary reference
point $\xi_1 = \tilde\xi$ to be defined later) by (1.2) in the main paper.
Note that, the empirical Bayes estimate of $\xi$
is the point where the marginal likelihood function is maximized
\cite[see e.g.][]{roy:evan:zhu:2016}. Since $\beta, \sigma^2$ can be
analytically integrated out, one can work with the posterior density
of $\bz$, $\pi_\xi(\bz | \by)$. Here, we consider multiple IS
estimator (1.2) in the main paper 
based on samples from the (transformed)
density $\pi_\xi(\bmu | \by)$ \citep[see][for the reasons for
considering the transformed samples]{evan:roy:2019}. Thus, here
$Q= \Pi = \{ \pi_\xi(\bmu | \by), \xi \in \Xi\}$ for some $\Xi$
defined later. 

In order to narrow down the potential region of search,
we initially choose a
wide range of values for each component of $\xi$, and form a large grid,
denoted by $\Xi$, by
combining discrete points within these ranges. This gives us the set
consisting of the following $9^4$ points:
\begin{equation*}
  \Xi = \{0, 0.5, \ldots, 4\} \times
  \{100, 425, \ldots, 2700\} \times \{0, 0.75,\ldots, 6\} \times \{0.1,
  0.35, \ldots, 2.1\}.
\end{equation*}
The SFE method, after each range is
scaled in $[0,1]$,  was applied to choose $k=5$ points from $\Xi$. Markov
chain samples
from the $k$ densities $\pi_{\xi_i}(\bmu | \by), i=1,\dots,k$ corresponding to this
preliminary skeleton set are
generated. We evaluate (1.2) in the main paper 
with
$N_l = 1000, n_l =1000, l=1,\dots, 5$ for all points in $\Xi$, and retain only
those points for which the value of (1.2) in the main paper 
is not less than 60\% of the maximum
value. The maximum value is attained at
$\tilde{\xi} = (1, 425, 2.25, 0.6)$ and there are 33 points
satisfying this criterion. These points form the search set
$\tilde\Xi$ which is a subset of
\begin{equation*}
  \tilde\Xi \subset \{1\} \times \{100, 425, \ldots, 1400\} \times
  \{1.50, 2.25, \ldots, 4.50\} \times \{0.35, 0.60, \ldots, 2.10\} .
\end{equation*}
Our aim is to choose $k=5$ elements from $\tilde\Xi$, one of which
must be $\tilde{\xi}$, to form the skeleton sets, using the
methods discussed in Section 3 of the main paper, 
with the objective of
estimating the ratios of marginal densities in $\Xi$ relative to
$\tilde{\xi}$. The naive IS method with samples from the posterior density
$\pi_{\tilde\xi}(\bmu | \by)$ is considered for comparison.

The SFE optimal set is computed on $\tilde\Xi$ after each dimension is
scaled in $[0,1]$. For SFS, we write (3.7b) in the main paper 
as an integral
over $(\bz,\log\sigma^2)$, because the prior for $\bz$ is multivariate
normal and $\sigma^2>0$, and use the approximation given in Section~\ref{sec:laplace}
of the supplementary materials. The SEQ, MNX, and ENT optimal sets are
computed iteratively. At each iteration, an estimate of the asymptotic
relative standard error is computed using Theorem 2(a) 
based on
1000 samples for the first stage and 1000 samples for the second
stage. We use different empirical convergence diagnostics
\citep{roy:2020} to check mixing of the Markov chains. Further
computational details about our implementation are provided in
Section~\ref{sec:compdetronge} 
of this supplementary materials.

\begin{figure}
  \centering
  \includegraphics[width=\linewidth]{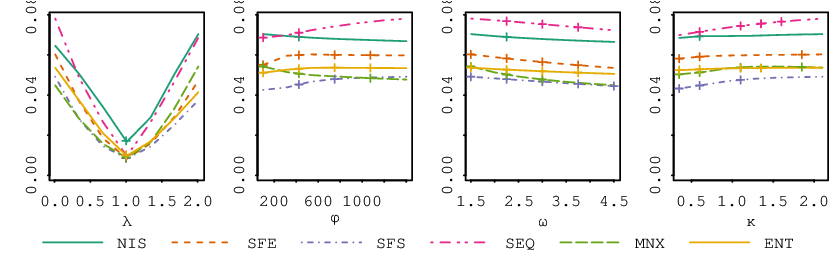}
  \caption{Profile relative standard errors estimates for the Rongelap data. One parameter
    is fixed and the maximum relative standard error across the other parameters is plotted
    against the fixed parameter. The crosses indicate points included
    in the skeleton set.}
  \label{fig:rongelap_se}
\end{figure}

Finally, for each obtained
skeleton set, we generate new samples which we use to estimate the
ratio of the marginal likelihoods and its relative standard error for all $\xi \in \Xi$. We generate a total of 50,000 samples
which are equally divided between each proposal density (see
Section~\ref{sec:compdetronge} for further details). 
The maximum relative standard error
estimates corresponding to one component of $\xi$ fixed across the
other components are shown in Figure~\ref{fig:rongelap_se}. It can be
seen that across all parameters, naive IS and SEQ have the
highest variance, and that SFS, MNX, ENT have the lowest maximum
variance.

\subsection{Computational details}
\label{sec:compdetronge}

Here we give more details on how we produced the results in this Section. 
The computation is done in two phases. In the
first phase we find the optimal skeleton set, $\bxi$, for each method. In the second
phase we compare the relative standard error of the naive and multiple IS estimators
using the skeleton sets computed in the first phase. The total number of
samples used for each method in the second phase is kept fixed and the same. In all
calculations involving the asymptotic variance of the IS estimator, the SV
estimate with the Tukey-Hanning window,
\[
  w_{n} (j) = 0.5[1 + \cos (\pi |j|/ b_{n})]I(|j| < b_{n}),
\]
was used, where $b_n = \sqrt{n}$. Markov chain samples are generated using
the algorithm of \cite{digg:tawn:moye:1998}.


\noindent{\bf Phase 1: Finding the optimal set of proposal densities}

\begin{itemize}
\item \textbf{NIS:} This is not required because the proposal density is
  fixed at $\bxi_\mathrm{nis} = \{\tilde\xi\} = \{(1,425,2.25,0.6)\}$.
\item \textbf{SFE:} This method is based on the Euclidean distance
  between the parameters. Each component of the parameter set $\tilde\Xi$ is
  scaled to vary between 0 and 1 before the optimal set is
  computed. The point-swapping algorithm of
  Section~\ref{sec:point-swapp-algor} is used to find the optimal
  set. We find\\
  $\bxi_\mathrm{sfe} = 
  \{\tilde\xi,(1,100,3,1.85),(1,425,3.75,0.6),(1,750,3,0.35),(1,1075,1.5,0.35)\}$.
\item \textbf{SFS:} This method requires the SKLD between two
  densities corresponding to two values of $\xi$. We first write the
  integrals in~(3.7a) of the main paper in terms of
  $(\bz,\log\sigma^2)$, because the prior for $\bz$ is multivariate
  normal and $\sigma^2>0$. Each ratio of integrals is approximated
  using the Laplace's method given in Section~\ref{sec:laplace}. The
  point-swapping algorithm of Section~\ref{sec:point-swapp-algor} is
  used to find the optimal
  set. We find\\
  $\bxi_\mathrm{sfs} =
  \{\tilde\xi,(1,425,3,0.6),(1,425,3.75,1.1),(1,425,4.5,0.6),(1,750,1.5,0.35)\}$.
\item \textbf{SEQ:} In this case the optimal set is computed by starting at
  $\bxi^{(1)} = \bxi_\mathrm{nis}$. Then, given that at the $i$th
  iteration, $i=2,\ldots,k$ we are at $\bxi^{(i-1)}$, we obtain $\bxi^{(i)}
  = \bxi^{(i-1)} \cup \{\xi'\}$, where $\xi'$ corresponds to the point in
  $\tilde\Xi\setminus\bxi^{(i-1)}$ with the highest relative standard error. The relative standard error
  is again computed using 1,000 Gibbs samples for stage~1 and 1,000
  new samples for stage~2. We find\\
  $\bxi_\mathrm{seq} = 
  \{\tilde\xi,(1,100,2.25,1.35),(1,100,2.25,1.6),(1,425,3,1.1),(1,425,3.75,1.1)\}$.
\item \textbf{MNX:} The optimal set is found by simulated annealing
  (Section~\ref{sec:simul-anne-algor}). We start the simulated annealing
  algorithm at $\bxi_\mathrm{sfs}$ and perform $i_\mathrm{max}=250$
  iterations with $T_0=0.001$ and $B=10$. 
  The SV
  estimates were calculated using 1,000 Gibbs samples for stage~1 and 1,000
  new samples for stage~2. We find\\
  $\bxi_\mathrm{mnx} = 
  \{\tilde\xi,(1,100,2.25,1.85),(1,425,1.5,0.6),(1,425,2.25,1.1),(1,1075,3,0.35)\}$.
\item \textbf{ENT:} The optimal set is found by simulated annealing
  (Section~\ref{sec:simul-anne-algor}). We start the simulated annealing
  algorithm at $\bxi_\mathrm{sfs}$ and perform $i_\mathrm{max}=250$
  iterations with $T_0=10$ and $B=10$. The optimality criterion used for
  this method is $-\log\mbox{det}(U)$ where $U$ is the matrix with $(i,j)$th
  element $U_{ij} = (\widehat{V}_{\bq})_{ij}/(\hat{d}_i\hat{d}_j)$. The
  estimates $\hat{\bd}$ and $\widehat{V}_{\bq}$ are computed from 1,000
  Gibbs samples. We find\\
  $\bxi_\mathrm{ent} = 
  \{\tilde\xi,(1,100,2.25,1.35),(1,425,3,1.1),(1,750,3.75,0.35),(1,1075,1.5,0.35)\}$.
\end{itemize}


\noindent{\bf Phase 2: Estimation of the ratio of marginal densities}
  
After the skeleton sets are found, we generate a total of $M=\text{50,000}$ samples
from the proposal densities, equally divided among all densities in the
set. Thus, for NIS we simply take $M$ samples from the density
corresponding to $\bxi_\mathrm{nis}$ and for the multiple IS methods we
take $M/k$ samples from each density in the corresponding $\bxi$ set. However, the
total of $M/k$ samples must be split into stage~1, for estimating the ratio
of marginals within $\bxi$, $\bd$, and stage~2, for estimating the
ratio of marginals over the whole set $\Xi$. To determine the optimal split
we use equation~\eqref{eq:relse} where $\hat\upsilon_1(\xi)$ and
$\hat\upsilon_2(\xi)$ are calculated from 1,000 stage~1 and 1,000 stage~2
samples. For MNX, SEQ, and ENT the existing samples from Phase~1 are
reused but for the space-filling methods, new samples are generated. We
take $\hat N/k$ stage~1 samples from each density in the skeleton set where
$\hat N$
is the integer that minimizes $\max_{\xi \in \Xi} \mathrm{RelSE}(\xi,N,M-N)$,
and $\hat n/k$ stage~2 samples, where $\hat n = M - \hat N$. This corresponds to stage~1 sample sizes of 500 and stage~2 sample sizes of 9,500
from each density for all methods. 


  \bibliographystyle{apalike}
\bibliography{ref1}
\end{document}